\documentclass[amsmath,amssymb,aps,pra,preprint,groupedaddress]{revtex4-1}

\usepackage[dvipdfmx]{graphicx}
\usepackage{dcolumn}
\usepackage{bm}
\usepackage{longtable}
\usepackage{ulem,soul}
\usepackage[dvipsnames]{xcolor}

\begin{document}


\title{The stability of 3D skyrmions under mechanical stress studied via Monte Carlo calculations }


\author{Sahbi El Hog$^{1}$}
\author{Fumitake Kato$^{2}$ }
\author{Satoshi Hongo$^{3}$ }
\author{Hiroshi Koibuchi$^{2}$}
\email[]{koibuchi@gm.ibaraki-ct.ac.jp; koibuchih@gmail.com}
\author{Gildas Diguet$^{4}$ }
\author{Tetsuya Uchimoto$^{5,6}$ }
\author{Hung T. Diep$^{7}$}
\email[]{diep@cyu.fr} 

\affiliation{
  $^{1}$Laboratoire de la Mati${\grave{e}}$re Condens${\acute{e}}$e et des Nanosciences (LMCN), Universit${\acute{e}}$ de Monastir, D${\acute{e}}$partement de Physique, Facult${\acute{e}}$ des Sciences de Monastir, Avenue de l'Environnement, 5019 Monastir, Tunisia~\\
  $^{2}$National Institute of Technology (KOSEN), Ibaraki College, Nakane 866, Hitachinaka, Ibaraki
312-8508, Japan~\\
$^{3}$National Institute of Technology (KOSEN), Sendai College, 8 Nodayama, Medeshima-Shiote, Natori-shi,
Miyagi 981-1239, Japan~\\
$^{4}$Advanced Institute for Materials Research (AIMR), Tohoku University, 2-1-1 Katahira, Aoba-ku, Sendai 980-8577, Japan ~\\
$^{5}$Institute of Fluid Science (IFS), Tohoku University, 2-1-1 Katahira, Aoba-ku, Sendai 980-8577, Japan~\\
$^{6}$ElyTMax, CNRS-Universite de Lyon-Tohoku University, 2-1-1 Katahira, Aoba-ku, Sendai, Japan~\\
$^{7}$Laboratoire de Physique The${\acute{o}}$rique et Mod${\acute{e}}$lisation,
CY Cergy Paris University, CNRS, UMR 8089 2, Avenue Adolphe Chauvin,
95302 Cergy-Pontoise Cedex, France
 }




\begin{abstract}
Using Monte Carlo (MC) simulations, we study the skyrmion stability/instability as a response to uniaxial mechanical stresses. Skyrmions emerge in chiral magnetic materials as a stable spin configuration under external magnetic field $\vec{B}$ with the competition of ferromagnetic interaction and Dzyaloshinskii-Moriya interaction (DMI) at low temperature $T$. Skyrmion configurations are also known to be stable (unstable) under a compressive stress applied parallel (perpendicular) to $\vec{B}$. To understand the origin of such experimentally confirmed stability/instability, we use the Finsler geometry modeling technique with a new degree of freedom for strains, which plays an essential role in DMI being anisotropic. We find from MC data that the area of the skyrmion state on the $B$-$T$ phase diagram increases (decreases) depending on the direction of applied stresses, in agreement with reported experimental results. This change in the area of the skyrmion state indicates that skyrmions become more (less) stable if the tensile strain direction is parallel (perpendicular) to $\vec{B}$. From the numerical data in this paper, we find that  the so-called magneto-elastic effect is suitably implemented in the effective DMI theory with the strain degree of freedom without complex magneto-elastic coupling terms for chiral magnetic materials. This result confirms that experimentally-observed skyrmion stability and instability are caused by DMI anisotropy.

\end{abstract}

\maketitle


\section{Introduction}
\label{intro} 
Skyrmions are topologically stable spin configurations in chiral magnetic materials such as MnSi, FeGe and 
${\rm Cu_2OSeO_3}$  and attract much attention for future spintronics devices \cite{Skyrme-1961,Moriya-1960,Dzyalo-1964,Romming-etal-Science2013,Fert-etal-NatReview2017,Zhang-etal-JPhys2020,Gobel-etal-PhysRep2021}. Studies conducted theoretically and experimentally have clarified that external magnetic fields and mechanical stresses play a crucial role in influencing skyrmion configurations \cite{Bogdanov-PRL2001,Bogdanov-Nat2006,Bogdanov-PHYSB2005,Bogdanov-SovJETP1989,Uchida-etal-SCI2006,Yu-etal-Nature2010,Butenko-etal-PRB2010}. Shape deformation of skyrmions by uniaxial stress has been experimentally studied, and it has been reported that the origin of the deformation is due to a direction-dependent Dzyaloshinskii-Moriya interaction (DMI) \cite{Moriya-1960,Dzyalo-1964}, in which a spin-orbit coupling is originally implemented \cite{Shibata-etal-Natnanotech2015,Koretsune-etal-SCRep2015}. The same conclusion on the origin has also been obtained numerically by the Finsler geometry (FG) modeling technique \cite{El-Hog-etal-PRB2021}. 

\begin{figure}[h!]
\centering{}\includegraphics[width=12.5cm]{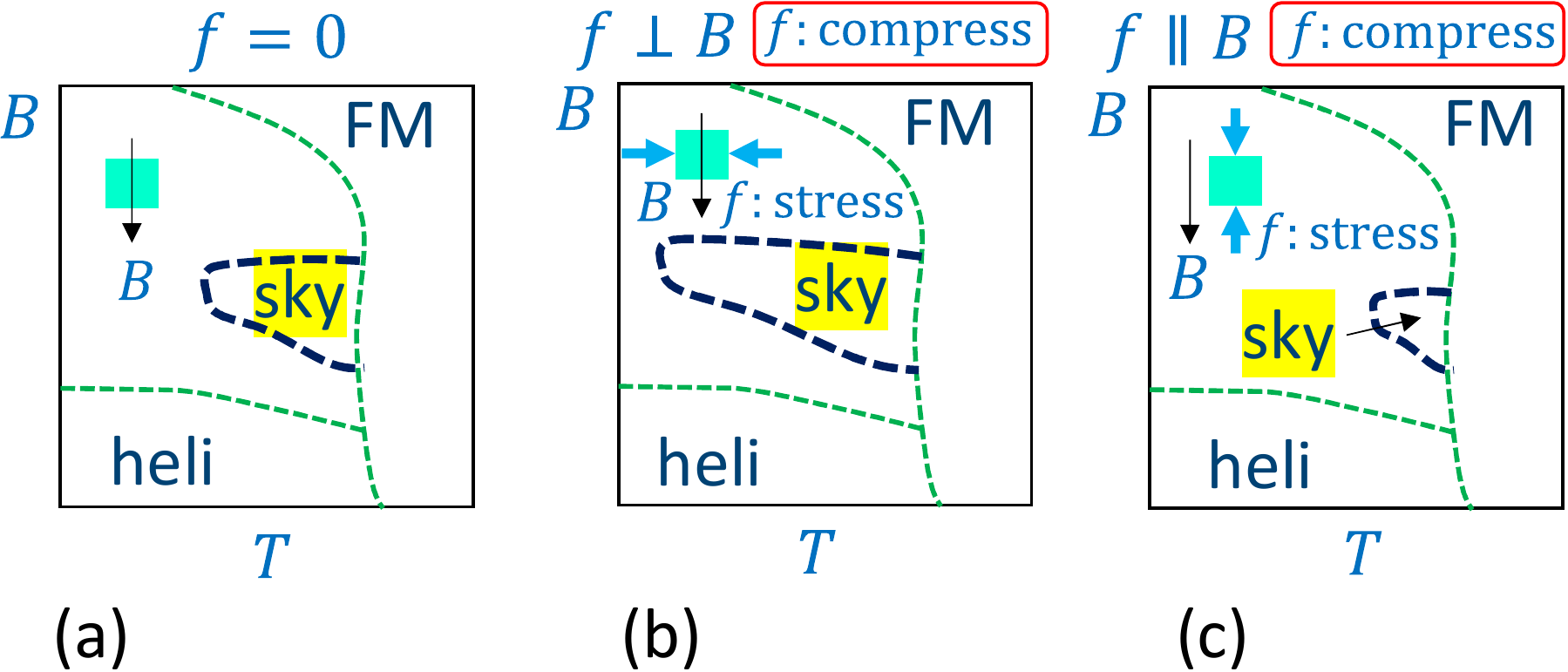}
\caption{
Illustrations of the change of skyrmion state area in the $B$-$T$ phase diagrams reported in Refs. \cite{Pfleiderer-etal-Science2009,Nii-etal-NatCom2015, Charcon-etal-PRL2015,Seki-etal-PRB2017}. The symbol $B$ is an applied magnetic field along $Z$ direction ($\Leftrightarrow \vec{B}\!=\!(0,0,-B)$), and $T$ is the temperature.  Compressive stress $f$ is zero in (a) $f\!=\!0$,  and the direction of $\vec{f}$  is (b) $\vec{f}\!\perp\! \vec{B}$  and (c) $\vec{f}\!\parallel \! \vec{B}$.  sky, heli and FM denote skyrmion, helical and ferromagnetic phases, and the dashed lines roughly represent the phase boundaries. 
}
\label{fig-1} 
\end{figure}
Uniaxial mechanical stresses can also stabilize skyrmion configurations \cite{Butenko-etal-PRB2010}, and experimental studies have reported that the uniaxial mechanical stresses stabilize or destabilize skyrmions depending on whether the compression direction is parallel or perpendicular to the magnetic field $B$   \cite{Pfleiderer-etal-Science2009,Nii-etal-NatCom2015, Charcon-etal-PRL2015,Seki-etal-PRB2017}. 
Recently, Wang et al. reported an electric-field-driven manipulation technique of individual skyrmions based on magneto-mechanical effects \cite{Wang-etal-NatCom2020}.

To make clear the problem, we illustrate a summary of some parts of corresponding reported experimental data in Figs. \ref{fig-1}(a)--(c)), where no detailed information on crystalline axes is given but only directions of stress and magnetic field and its relation are specified. This phenomenon has also been studied in terms of magneto-elastic coupling (MEC) effects, and  precise calculations have been made using Landau-type theories \cite{Shi-Wang-PRB2018,Wang-Shi-Kamlah-PRB2018,Wang-ARMR2019}.  In such a phenomenological theory,  DMI is not directly deformed,  and instead, an explicit MEC is assumed in the free energies at the beginning.  Therefore, the observed skyrmion stability/instability is the consequence of such an assumption. 

MEC is known to stem from complex Coulomb interactions between atoms and electrons. The standard modeling technique to include such MEC in the exchange or ferromagnetic interaction (FMI)  assumes that FMI depends on the distance between spins. Hence, such a phenomenological model, which includes the distance variable, allows for a precise calculation in the assumed range of distance in chiral magnetic materials \cite{Kataoka-JPSJ1974,Kataoka-JPSJ1987}. However, the Landau-type model is complex, especially for anisotropic phenomena, due to some additional terms representing direction-dependent MEC contributions \cite{Plumer-etal-JPC1984,Plumer-Walker-JPC1982}. 

Now, if the coefficient of FMI or DMI dynamically depends on strains, a different story is possible. In such a case of anisotropic interactions, magneto-elastic effects on skyrmions can be studied in the framework of effective interaction theories of FMI and DMI without additional terms for MEC in the free energy, in contrast to the case of Landau-type models as described above. In the FG model, MEC is not assumed as input but obtained as an output. The only problem that should be clarified is whether such an approach based on the theory of effective interactions with dynamical anisotropy is meaningful or not. This validity should be checked carefully by comparing the numerical results with reported experimental ones in \cite{Pfleiderer-etal-Science2009,Nii-etal-NatCom2015,Charcon-etal-PRL2015,Seki-etal-PRB2017}. 

In this paper, we show that the DMI anisotropy reproduces skyrmion stability/instability as experimentally reported. In the FG model,  a strain field $\vec{\tau}(\in S^2/2: {\rm half \; sphere},  S^2: {\rm sphere})$ is introduced to cause a DMI anisotropy, and no MEC term is needed. This 3D FG model is an almost straight-forward extension of the 2D model in Ref. \cite{El-Hog-etal-PRB2021}.  Our conclusion is that the dynamically implemented magneto-elastic effect causes DMI anisotropy, and the DMI anisotropy causes skyrmion stability/instability consistently with reported experimental data. This mechanism, pointed out in Ref. \cite{Seki-etal-PRB2017},  is exactly the same as in the 2D model of Ref. \cite{El-Hog-etal-PRB2021} for 2-dimensional anisotropy, though 3D simulations are slightly more complex and time consuming compared with those of the 2D model.

A recent numerical study precisely reproduced experimental data of stability/instability by assuming as input an anisotropy in the DMI coupling coefficients \cite{Tanaka-etal-PRM2020}, while in  our model, this anisotropy in the DMI comes from the strain acting on the DMI. It is not surprising that in Ref. \cite{Tanaka-etal-PRM2020},  the area of skyrmion state is almost exactly identical to the experimental one by tuning the DMI anisotropy, while in our simulations using a thin 3D disk, the DMI anisotropy comes out as a consequence of a strain application. Our results, however, are consistent with those of Ref. \cite{Tanaka-etal-PRM2020} and with the results obtained for the bulk 3D materials \cite{Yu-etal-PRB2015}.  Note that we have two types of theoretical models for DMI, one is bulk DMI, and the other is interfacial DMI, in which Bloch type and Neel type skyrmions are expected, respectively. In
this paper, we concentrate on the bulk DMI, although almost the same results are expected in the case of interfacial DMI.

\section{Models}
\label{models} 

\subsection{The Hamiltonian and a new variable for mechanical strains}
\label{Hamiltonian-partition-function}
We use a 3D disk composed of tetrahedrons to obtain a discrete Hamiltonian (see Appendix \ref{3D-disk}). 
The discrete Hamiltonian is given by the linear combination of five terms such that 
\begin{eqnarray}
 \label{total-Hamiltonian}
S=\lambda S_{{\rm FM}}+DS_{{\rm DM}}-S_{B}+\gamma S_{\tau}-S_{f},
\end{eqnarray}
where FMI and DMI energies $S_{{\rm FM}}$ and $S_{{\rm DM}}$ are
given in two different combinations denoted by model 1 and model 2, as in Ref. \cite{El-Hog-etal-PRB2021}
\begin{eqnarray}
\label{model-1}
S_{{\rm FM}}=\sum_{\it \Delta}\sum_{ij({\it \Delta})}\bar{\Gamma}_{ij}\left(1-\vec{\sigma}_{i}\cdot\vec{\sigma}_{j}\right),\quad 
 S_{{\rm DM}}=\sum_{ij}{\vec{e}}_{ij}\cdot\vec{\sigma}_{i}\times\vec{\sigma}_{j}, \quad({\rm model\;1}),
\end{eqnarray}
and 
\begin{eqnarray}
\label{model-2}
S_{{\rm FM}}=\sum_{ij}\left(1-\vec{\sigma}_{i}\cdot\vec{\sigma}_{j}\right),\quad 
 S_{{\rm DM}}=\sum_{\it \Delta}\sum_{ij({\it \Delta})}\bar{\Gamma}_{ij}{\vec{e}}_{ij}\cdot\vec{\sigma}_{i}\times\vec{\sigma}_{j}, \quad({\rm model\;2}).
\end{eqnarray}
The discrete form of these Hamiltonians and the definition of the effective coupling constant $\bar{\Gamma}_{ij}$ are described in Appendix \ref{discrete-FMDM}. 

Note that $\bar{\Gamma}_{ij}$ depends on the strain variable $\vec{\tau}$, which will be described below, and this  $\bar{\Gamma}_{ij}$ is defined to be $\bar{\Gamma}_{ij}\to 1$ in the isotropic case, i.e., randomly distributed, $\vec{\tau}$. As a consequence, $S_{\rm FM}$ in model 1 and  $S_{\rm DM}$ in model 2 reduce to the standard  $S_{\rm FM}$  and $S_{\rm DM}$ for the isotropic case.  More detailed information is given in Appendix \ref{discrete-FMDM}.

Another possible interaction stabilizing skyrmions is the dipole-dipole interaction. However, this term is known to be effective for objects of which the size is very large up to several micrometers in diameter while
skyrmions created by DMI or by other frustrated short-range interactions have sizes up to a few hundred nanometers in diameter \cite{Gobel-etal-PhysRep2021}. For this reason,
the dipole-dipole interaction is neglected in this work.

$S_{\rm DM}$ in model 1 and $S_{\rm FM}$ in model 2 are the standard ones, where interactions are direction-independent or isotropic, and $\sum_{ij}$ denotes the sum over bonds $ij$. On the contrary, anisotropic  interactions are  implemented in $S_{{\rm FM}}$ of model 1 and $S_{{\rm DM}}$ of model 2, where $\sum_{\it \Delta}$ and $\sum_{ij({\it \Delta})}$ denote the sum over tetrahedrons ${\it \Delta}$ and the sum over 6 bonds $ij({\it \Delta})$ of the tetrahedron ${\it \Delta}$, respectively. The spin variable at vertex $i$ is denoted by $\vec{\sigma}_i$, which has values on the unit sphere; $\vec{\sigma}_i \in S^2$. The parameters $\lambda$ and $D$ in Eq. (\ref{total-Hamiltonian}) are the coupling constants of FMI and DMI, respectively. Note that $\lambda \bar{\Gamma}_{ij}$ of $S_{\rm FM}$ in model 1 is the effective coupling constant for FMI. This coupling constant is not constant but depends on position and direction, and $D \bar{\Gamma}_{ij}$ of $S_{\rm DM}$ in model 2 is also the position-dependent and direction-dependent coupling constant for DMI. 
In Appendix \ref{E-representation}, we present detailed information on how mechanical strains influence FMI and DMI energies and dynamically deform their coefficients to anisotropic, i.e., how anisotropic interactions are implemented. We show that only model 2 is suitable for the reported experimental data.

\begin{figure}[t]
\centering{}\includegraphics[width=8.5cm]{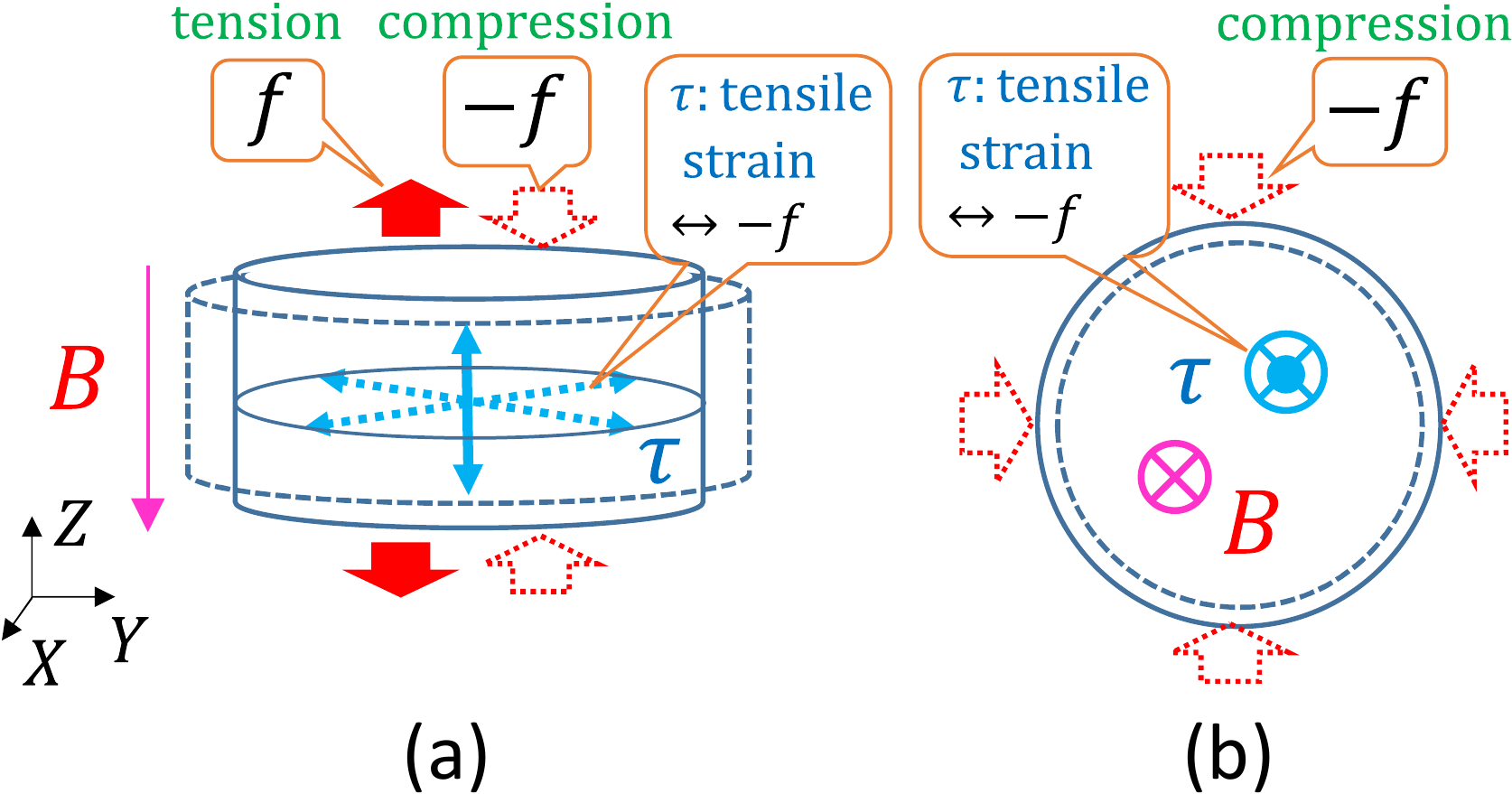}
\caption{(a) Illustrations of the 3D disk,  stresses $\vec{f}\!=\!(0,0,\pm f)$, strains $\vec{\tau}$ and magnetic field $\vec{B}\!=\!(0,0,-B)$, (b) the directions of $\vec{\tau}$  and $\vec{B}\!=\!(0,0,-B)$ are parallel to $Z$ direction under the lateral pressure. This direction of $\vec{\tau}$ in (b) is the same as in (a) for the tensile $f$.  In the simulations,  $f$ or $-f$ is applied to the direction parallel to $\vec{B}$, as in (a). The dashed lines of the disk shape in (a) and (b) are drawn to show the expected shape changes in real materials, while the changes in volume and shape are neglected and remain unchanged in the simulations. 
}
\label{fig-2} 
\end{figure}

The other terms in Eq. (\ref{total-Hamiltonian})  are given by 
\begin{eqnarray}
\begin{array}{l}
\label{other-terms}
 S_{B}=\sum_{i}\sigma_{i}\cdot\vec{B},\quad\vec{B}=(0,0,B), \quad
 S_{\tau}=\frac{1}{2}\sum_{ij}\left(1-3(\vec{\tau}_{i}\cdot\vec{\tau}_{j})^{2}\right),
 \\
 S_{f}={\rm sgn}(f)\sum_{i}\left(\vec{\tau}_{i}\cdot\vec{f}\right)^{2},\quad{\vec{f}}=(f_x,f_y,f_z), 
 \quad {\rm sgn}(f)=\left\{ \begin{array}{@{\,}ll}
                 1 &  ({\rm tension}) \\
                 -1&  ({\rm compression}) 
                  \end{array} 
                   \right.,   \\            
\end{array}
\end{eqnarray}
In the Zeeman energy $S_B$, magnetic field $\vec{B}$ is in $Z$ direction, which is perpendicular to the disk (Fig. \ref{fig-2}(a)). The final two terms are introduced for the strain variable $\vec{\tau}_i$, which has values on the half sphere $\vec{\tau}_i\in S^2/2$ because $\vec{\tau}$ is assumed to be non-polar like the director field of liquid crystal molecules \cite{Doi-Edwards-1986}. 
The vector $\vec{f}$ in $S_f$ denotes an external mechanical force or stress, which is tensile or compressive. Because of the definition of $S_f$, the direction of $\vec{\tau}$ becomes parallel or perpendicular to $\vec{f}$ (Fig. \ref{fig-2}(a)). The direction of $\vec{\tau}$ becomes parallel to the tensile force direction, while it will be perpendicular to the compressive stress direction. Suppose the stress is lateral and compressive (Fig. \ref{fig-2}(b)), which is not uniaxial, then $\vec{\tau}$ is directed to the  $Z$  direction, which is parallel to $B$. Therefore, we numerically check whether these two different stresses, tensile and compressive(Fig. \ref{fig-2}(a)), stabilize or destabilize skyrmion configurations. The parameter $\gamma$ is fixed to a small non-zero value in the simulations to make $\vec{\tau}$ easily align along the direction of $\vec{f}$.
We should note that the strain variable $\vec{\tau}$ is introduced to be parallel (perpendicular) to tensile (compressive) stress directions, and this definition is different from the standard definition of mechanical stress in Refs.  \cite{Pfleiderer-etal-Science2009,Nii-etal-NatCom2015,Charcon-etal-PRL2015,Seki-etal-PRB2017}, where the stress $\sigma$ is introduced as a variable parallel to the compressive directions.

\noindent
\subsection{Monte Carlo technique and snapshots\label{MC-technique}}
The partition function to be simulated by Metroplis Monte Carlo (MC)  technique \cite {Metropolis-JCP-1953,Landau-PRB1976} is
\begin{eqnarray}
\label{part-func}
Z=\sum_{\vec{\sigma}}\sum_{\vec{\tau}}\exp\left[-S(\vec{\sigma},\vec{\tau})/T\right],
\end{eqnarray}
where $\sum_{\vec{\sigma}}$ and $\sum_{\vec{\tau}}$ denote the sum over all possible configurations of $\vec{\sigma}$ and $\vec{\tau}$. 
Both $\vec{\sigma}$ and $\vec{\tau}$ are randomly generated independent of the previous values on the unit sphere $S^2$ and accepted with the probability ${\rm Max}[1, \exp(-\delta S)]$, where $\delta S\!=\!S_{\rm new}\!-\!S_{\rm old}$. We should note that the interaction energies $S_\tau$ and $S_f$ in Eq. (\ref{other-terms}) are quadratic with respect to $\vec{\tau}$, and hence, $\vec{\tau}$ and $-\vec{\tau}$ in $S^2$ are identified to have a value on the half sphere $S^2/2$.

\noindent
\begin{longtable}{ccp{83mm}}
\caption{List of symbols and descriptions of the input parameters.  }
  \label{table-1} \\
  \hline
Symbol   & & Description  \\ \hline
\endfirsthead
\multicolumn{3}{r}{continue} \\ \hline
Symbol  & & Description (assumed typical value)  \\ \hline
\endhead
  \hline
\multicolumn{3}{r}{continue} \\
\endfoot
  \hline
\multicolumn{3}{r}{} \\
\endlastfoot
$T$  & & Temperature   \\
$\lambda$ & &   Ferromagnetic interaction coefficient    \\
$D$ & &   Dzaloshinskii-Moriya interaction coefficient    \\
$B$ & &   Magnetic filed    ${\vec B}=(0,0,-B)$ with $B\!>\!0$ \\
$\gamma$ & &  Interaction coefficient of $S_\tau$   \\
$f$ & &  Force strength  ${\vec f}=(0,0,f)$ \\
${\rm sgn}(f)$ & &  ${\rm sgn}(f)\!=\!1$ (tension), \quad   ${\rm sgn}(f)\!=\!-1$ (compression) \\
$v_0$ & &   Strength of anisotropy, which is fixed to $v_0\!=\!0.1$ for both FMI and DMI    \\
$\xi$ & &   Deformation parameter for the disk shape: $\xi\!=\!1 \Leftrightarrow$ non-deformed    \\
\end{longtable}
The parameters to be varied or fixed in the simulations are shown in Table \ref{table-1}.

\begin{figure}[h]
\centering{}\includegraphics[width=11.5cm,clip]{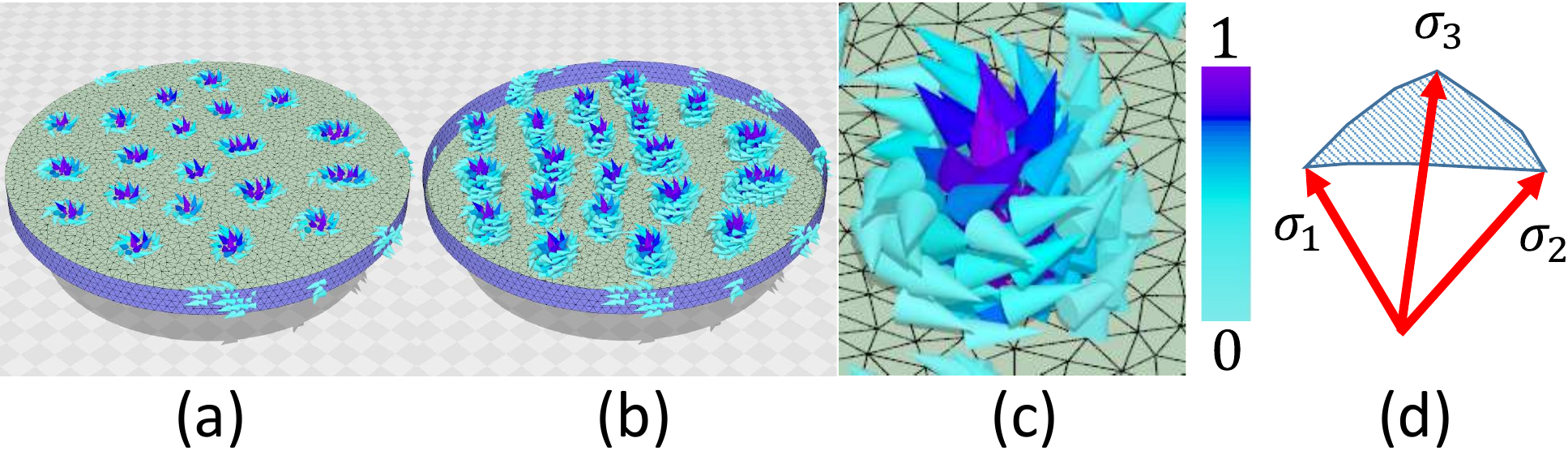}
\caption{(a) An upside view of a ground state configuration of skyrmions in the 3D disk, (b) the corresponding spins of $\sigma_z\!\geq\!0$ inside the disk with lower and side parts of the disk, (c)  a magnified skyrmion configuration, (d) a triangular area on the unit sphere defined by three spins of a triangle on the surface of the disk. The shaded area in (d) is used to calculate the total number of skyrmions. Note that skyrmions refer to not only spins of $\sigma_z\!\geq\!0$ visualized in (c) but also those of  $\sigma_z\!<\!0$ outside of them.  }
\label{fig-3} 
\end{figure}
Isotropic configurations or equivalently a random start is assumed for the initial configuration of $\vec{\sigma}$ and $\vec{\tau}$. The first step in MC simulation is to find a vacuum, i.e.  ground state  \cite{Hog-etal-JMagMat2018} for $\vec{\sigma}$ under the initial isotropic state of $\vec{\tau}$. Since $\vec{\tau}$ is isotropic, we expect that the ground state configuration of $\vec{\sigma}$ also becomes isotropic, independent of whether it is in the skyrmion (sky) phase or in other phases. 
Here, isotropic means ``not uniformly or globally deformed'', and for this reason, the skyrmion shape can locally deviate from the circular shape because of an expected local deformation of the DMI coefficient.

Figure \ref{fig-3}(a) shows an upside view of a ground state configuration in the skyrmion phase. For both $\vec{\sigma}$ and $\vec{\tau}$, the free boundary condition is assumed on all the boundary surfaces, including the circular edges of the 3D disk. For this ground state configuration, we show the spins of $\sigma_z\!\geq\!0$ in Fig. \ref{fig-3}(b), where the lower and the side parts of the disk are shown, and the other parts inside the disk are removed to make clear the spin configuration. This technique to visualize 3D configurations of spins is effective not only for the skyrmion phase but also for the other phases, such as the helical or stripe phases. Hence, this visualization technique is also used in the following section. From a single skyrmion configuration in Fig. \ref{fig-3}(c), we find that the skyrmion is a Bloch-type configuration. 

The reason why the free boundary condition is assumed is that the size of real materials
is finite.  In addition, the free surface condition is more suitable for non-uniform spin configurations as in the case of skyrmions, while the periodic boundary condition requires the sample size to be commensurate with the spin-angle periodicity which is not easy to detect. Note that the surface condition, free or periodic boundary condition, affects only spins up to a few atomic spacings from the surface so that the use of large samples will diminish the surface effect.  The result will not depend on the surface condition  if we use large sample sizes to reduce the surface effect.  Our choice is the free boundary condition with large samples.   In some specific cases, such as in Ref. \cite{Wang-etal-NatCom2020}, the material size is small enough so
that the periodic boundary condition is not suitable. In addition, we note that the present paper
does not focus on the so-called geometric confinement effect realized in some experiments. 

We should note that a conical state, of which the direction is along $\pm Z$ direction,  is expected in the 3D model for $\vec{B}\!=\!(0,0,-B)$ in the regions close to ($+Z$ direction) and far from  ($-Z$ direction)  the center of skyrmions. However, the thickness of the 3D disk is not sufficient to visualize the conical states, and for this reason, numerical data on the conical phase are not included, as in the 2D model in Ref. \cite{El-Hog-etal-PRB2021}.

The total number of MC sweeps is $1\times 10^6$ to find a ground state. We perform $5\times 10^7$ MC sweeps for every $(B,T)$ point starting from the ground state and draw the $B$-$T$ phase diagrams using the final configurations. The mean values of physical quantities are calculated from the configurations obtained every $1000$ MC sweeps by discarding the first $0.5\times 10^7$ MC sweeps for the thermalization. $0.5\times 10^7$ MC sweeps are sufficiently large for the thermalization inside the three different phases, where the convergence speed is fast.  At the phase boundaries,  the thermalization MC sweep is increased up to $4.5\times 10^7$ depending on the convergence behavior.

\section{Simulation results\label{results}}

\subsection{Shape deformation by uniaxial stresses \label{uniaxial-stress}}
\begin{figure}[h]
\begin{tabular}{cc}
\begin{minipage}{0.5\linewidth}
\centering{}\includegraphics[width=8cm]{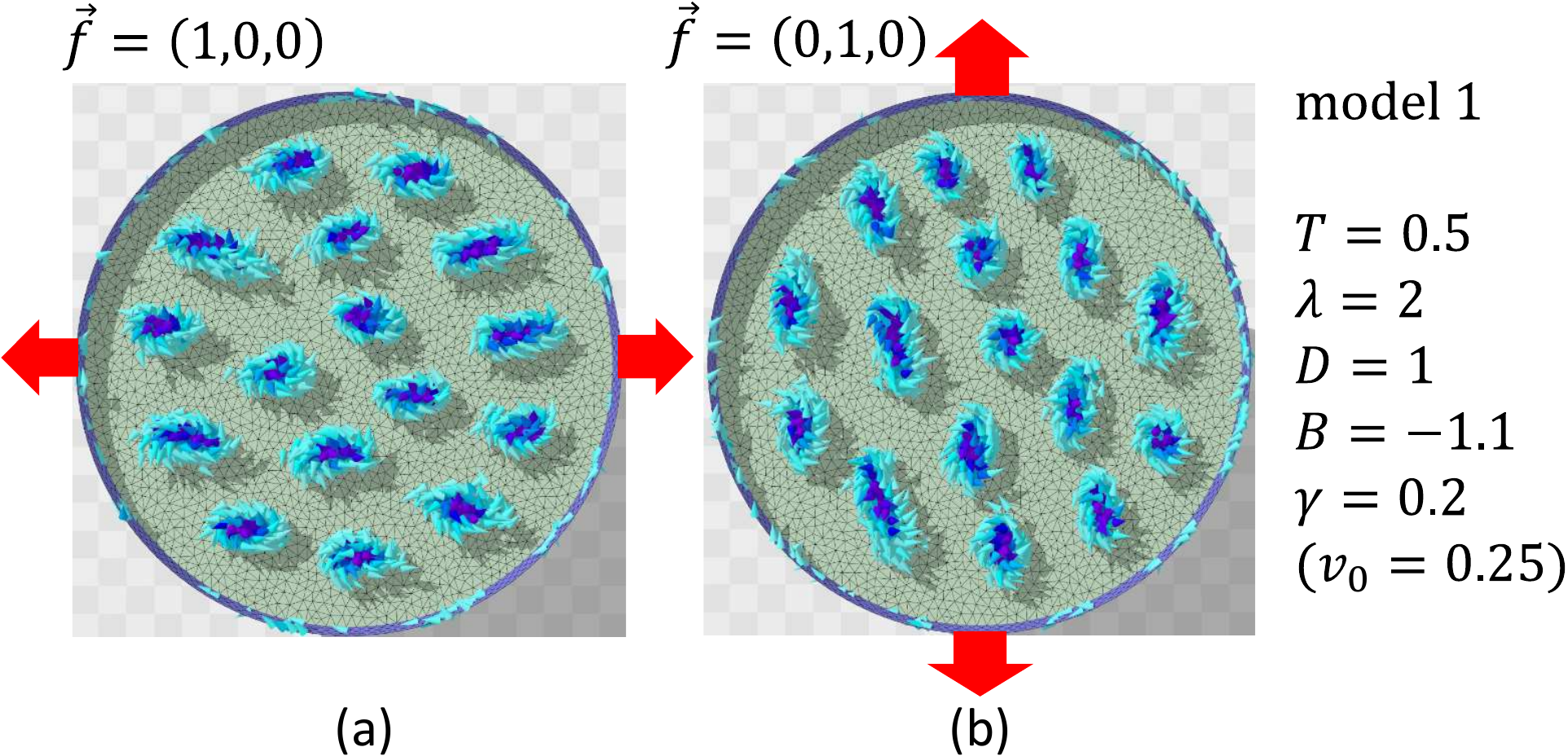} 
\caption {Snapshots of skyrmion shape deformation of model 1 under uniaxial tensile stresses of (a) $\vec{f}\!=\!(1,0,0)$  and (b) $\vec{f}\!=\!(0,1,0)$.  
\label{fig-4} }
\end{minipage}
\hspace{0.02\linewidth}
\begin{minipage}{0.5\linewidth}
\centering{}\includegraphics[width=8cm]{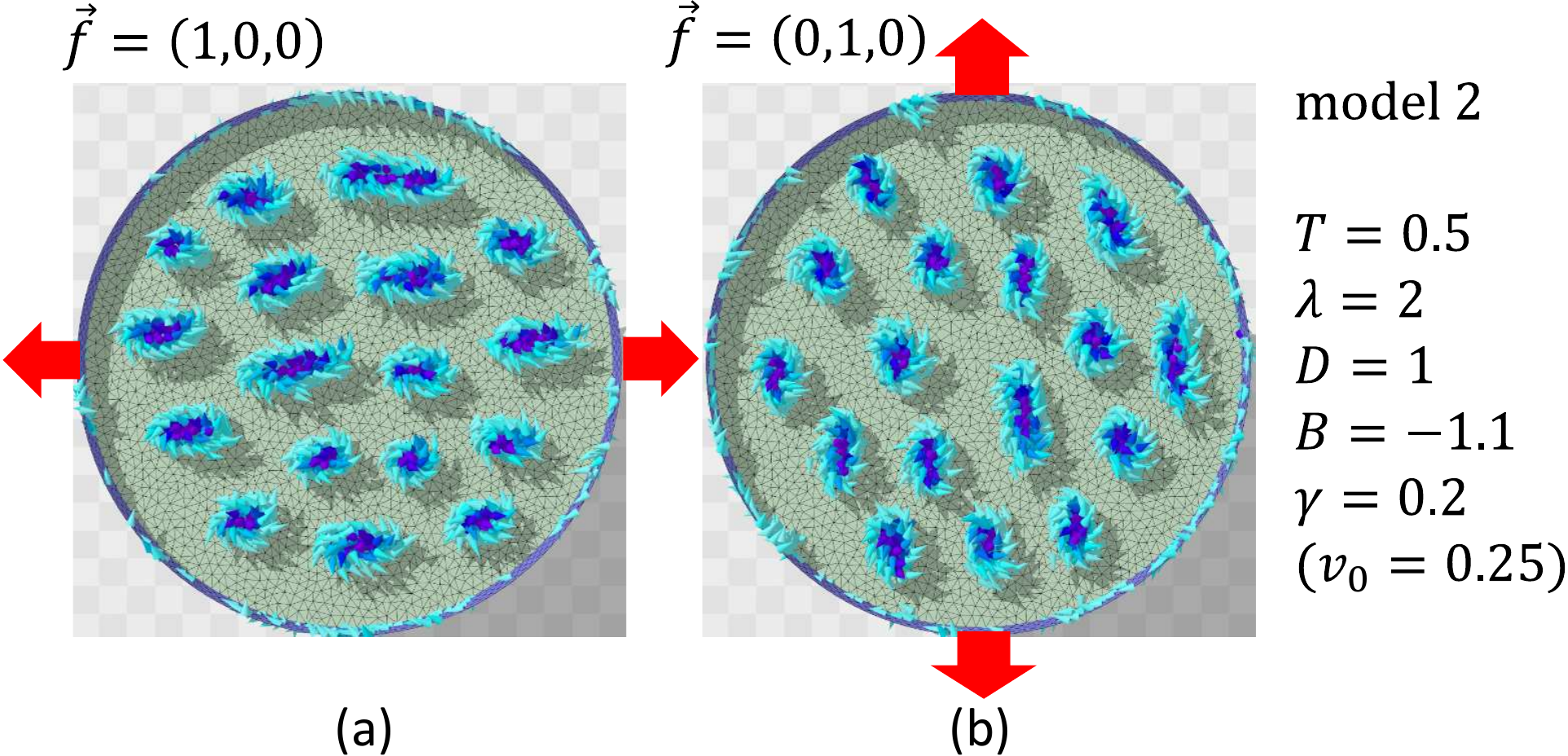}
\caption {Snapshots of skyrmion shape deformation of model 2 under uniaxial tensile stresses of (a) $\vec{f}\!=\!(1,0,0)$  and (b) $\vec{f}\!=\!(0,1,0)$.  
 \label{fig-5} }
\end{minipage}
\end{tabular}
\end{figure}
First, we show that the skyrmion shape deformation reported in \cite{Shibata-etal-Natnanotech2015} can be reproduced by both 3D model 1 and model 2, like in the 2D models in Ref. \cite{El-Hog-etal-PRB2021}. Snapshots of skyrmions under the tensile stresses $\vec{f}\!=\!(1,0,0)$ and $\vec{f}\!=\!(0,1,0)$ of model 1 and model 2 are shown in Figs. \ref{fig-4}(a),(b) and Figs. \ref{fig-5}(a),(b). The parameters assumed in the simulations, written on the figures, are the same in both models except the coefficients $(\lambda, D)$ of FMI and DMI. We find that the skyrmion shape deforms oblong in the tensile force directions in both models. This result is consistent with the experimental data in Ref. \cite{Shibata-etal-Natnanotech2015}, and it is also consistent with the numerical results of the 2D models in Ref. \cite{El-Hog-etal-PRB2021}. The shape anisotropy, controlled by $f$,  can be measured numerically on the upper surface using the same technique for a 2D lattice in Ref. \cite{El-Hog-etal-PRB2021}; however, we do not detail this problem.

Real strains corresponding to lattice deformations can also be assumed in the 3D disk, and we study responses to the real strains in the helical phase and present the results in Appendix \ref{uniaxial-strains}. We find that only model 2 reproduces stripe shapes, which is consistent with the reported experimental data.

Thus, we find at this stage that only model 2 is plausible as a model for the responses of skyrmion and helical phases to uniaxial stresses, as in the 2D case in Ref. \cite{El-Hog-etal-PRB2021}. The problem we are considering in this paper is whether the same conclusion can be drawn for the phenomenon of skyrmion stabilization/destabilization under uniaxial stresses. This is checked in the following subsections and the main purpose of this study.

\subsection{Stabilization and destabilization by uniaxial stresses \label{uniaxial-stress}}

\subsubsection{Magnetic field vs. Temperature diagram\label{BT-diagram}
}
\begin{figure}[h]
\begin{tabular}{cc}
\begin{minipage}{0.5\linewidth}
\centering{}\includegraphics[width=8cm]{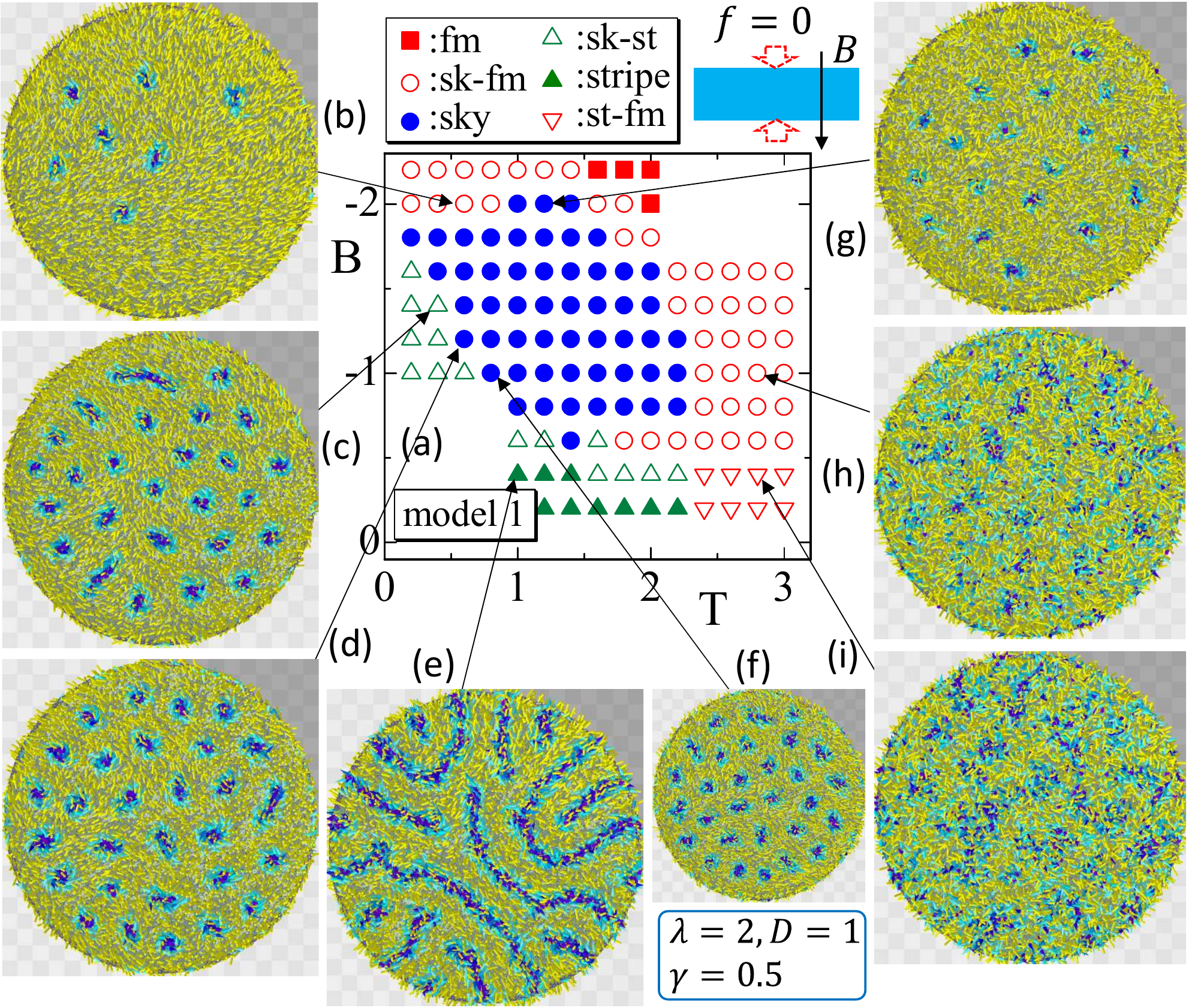} 
\caption {(a) $B$-$T$ phase diagram and (b)--(i)  snapshots of model 1 under $\vec{f}\!=\!(0,0,0)$.  Small cylinders (\textcolor{Dandelion}{\large{$ -$}} : yellow small cylinder) denote the directions of $\vec{\tau}$, which are almost (not always) random at large (small) $T$ because of the finite $\gamma\!=\!0.5$.
\label{fig-6} }
\end{minipage}
\hspace{0.02\linewidth}
\begin{minipage}{0.5\linewidth}
\centering{}\includegraphics[width=8cm]{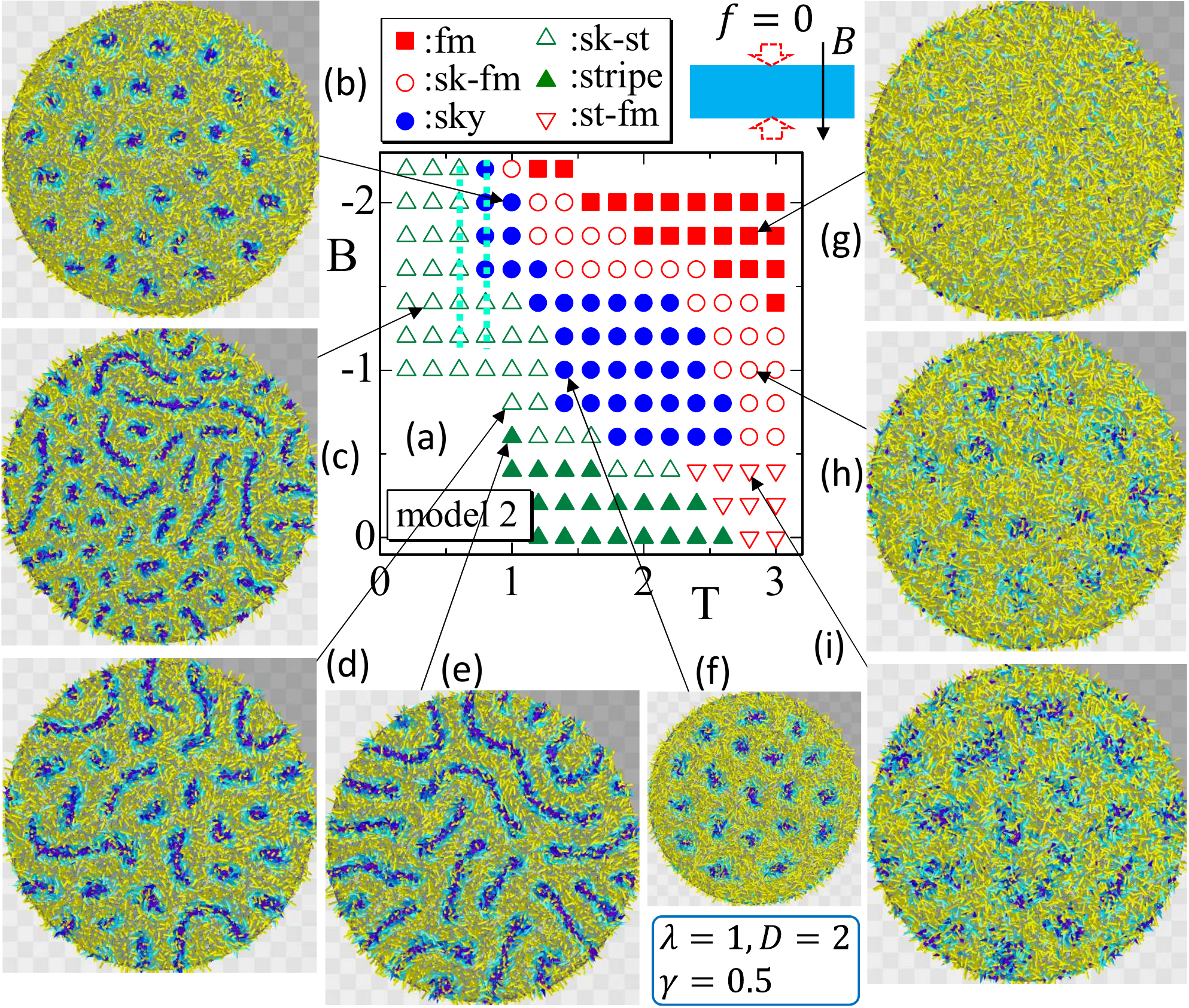} 
\caption {(a) $B$-$T$ phase diagram and  (b)--(i)  snapshots  of model 2 under $\vec{f}\!=\!(0,0,0)$. Small cylinders (\textcolor{Dandelion}{\large{$ -$}} : yellow small cylinder) denote the directions of $\vec{\tau}$, which are almost (not always) random at large (small) $T$ like in the case of model 1 in Fig. \ref{fig-6}.
\label{fig-7} }
\end{minipage}
\end{tabular}
\end{figure}
First, we show $B$-$T$ phase diagrams and snapshots of model 1 and model 2 obtained under zero stress $\vec{f}\!=\!(0,0,0)$ in Figs. \ref{fig-6} and \ref{fig-7}. The assumed parameters $\lambda$, $D$, and $\gamma$ are shown in the lower part of the figures. We assume that $(\lambda, D)\!=\!(2,1)$ for model 1 and $(\lambda, D)\!=\!(1,2)$ for model 2. The reason why $\lambda(=\!2)$ for model 1 is 2 times larger than $\lambda(=\!1)$ for model 2 is that  
the effective coupling constant $\bar{\Gamma}_{ij}$ in $S_{\rm FM}$ of model 1 is smaller than 1 (approximately $\bar{\Gamma}_{ij}\!\simeq\! 0.4\sim 0.7$), which is shown in the final part of this section. For the same reason, $D(=\!2)$ for $S_{\rm DM}$ in model 2 is 2 times larger than $D(=\!1)$ in model 1.

We observe three different phases, skyrmion (sky), stripe (stripe) and ferromagnetic (fm) phases, in the region of $0.2 \leq T\leq 3$ and $0.2\leq B\leq 2.2$. These are shown by solid symbols as depicted at the top of the figures. The empty symbols denote intermediate phases between two phases, denoted by sk-st, sk-fm, and st-fm, representing mixed phases of skymion and stripe, skyrmion and ferromagnetic, and stripe and ferromagnetic, respectively. We should note that the skyrmion-ferromagnetic phase changes to ferromagnetic at large $|B|$ ($|B|\!>\!2$ for model 1 and $|B|\!>\!1.5$ for model 2)  if the total number of MC sweeps is sufficiently large. The present results suffer from critical slowing down or the so-called trapping in a local minimum of energy because skyrmion is separated from ferromagnetic phase by a first-order phase transition in this region. However, the purpose of this study is to find a skyrmion region in the $B$-$T$ phase diagram, and we consider that the present scale of MC sweeps is sufficient for estimating the skyrmion region. Hence, long simulations are not always necessary at the phase boundaries. The paramagnetic phase is expected to appear in the region of higher temperature $T\!>\!3$ at $B\!\to\! 0$, however, in the region below $T\!=\!3$, the stripe still does not completely disappear.

From Figs. \ref{fig-6} and \ref{fig-7}, we find that skyrmions emerge  in almost the same region of the $B$-$T$ diagram for the case of $\vec{f}\!=\!(0,0,0)$ in both models. 
The areas of skyrmion state in both models are larger than the experimental data in \cite{Pfleiderer-etal-Science2009,Nii-etal-NatCom2015, Charcon-etal-PRL2015,Seki-etal-PRB2017}. However, we consider that our results are meaningful, because the thin 3D disk is used in the simulations and a stabilization is expected on such a thin-plate as reported in Ref. \cite{Yu-etal-PRB2015}.
 Note also that the skyrmion phase appears in the low $T$ region for the range of relatively large $|B|$ in model 1. Vertical dashed lines in Fig. \ref{fig-7}(a) denote the temperatures $T\!=\!0.6$ and $T\!=\!0.8$, at which $B$-$f$ phase diagrams are calculated to see stability/instability of skyrmions by varying $f$  in the following subsection.

\begin{figure}[h]
\begin{tabular}{cc}
\begin{minipage}{0.5\linewidth}
\centering{}\includegraphics[width=8cm]{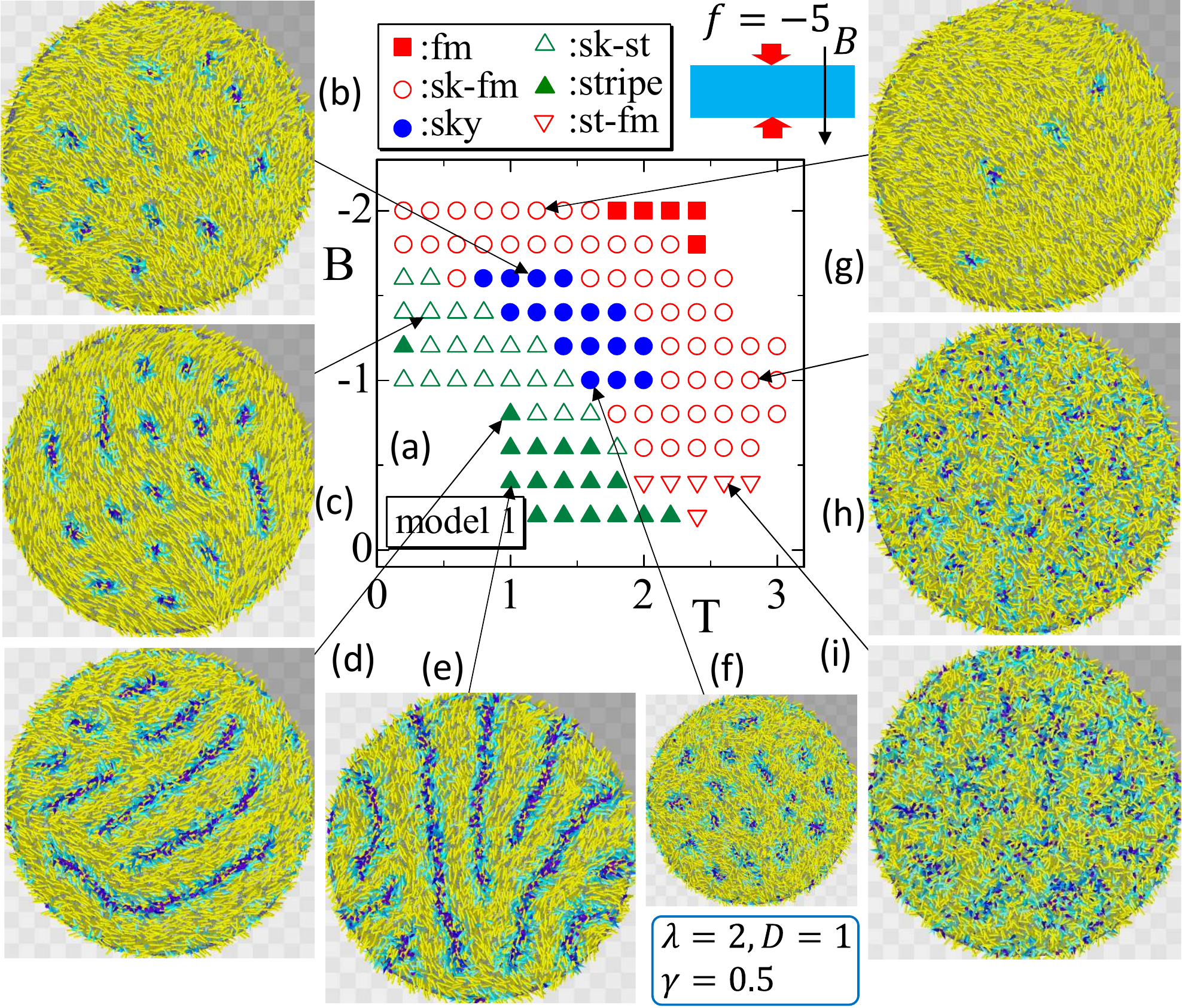} 
\caption {(a) $B$-$T$ phase diagram and (b)--(i)  snapshots of model 1 under compression $\vec{f}\!=\!(0,0,-5)$. The other parameters are the same as in Fig. \ref{fig-6}. The area of the skyrmion state shrinks compared with the case $\vec{f}\!=\!(0,0,0)$ in Fig. \ref{fig-6}(a). Strains $\vec{\tau}$ (\textcolor{Dandelion}{\large{$ -$}}) are perpendicular to $Z$ direction, i.e., in-plane directions, and visible.
\label{fig-8}}
\end{minipage}
\hspace{0.02\linewidth}
\begin{minipage}{0.5\linewidth}
\centering{}\includegraphics[width=8cm]{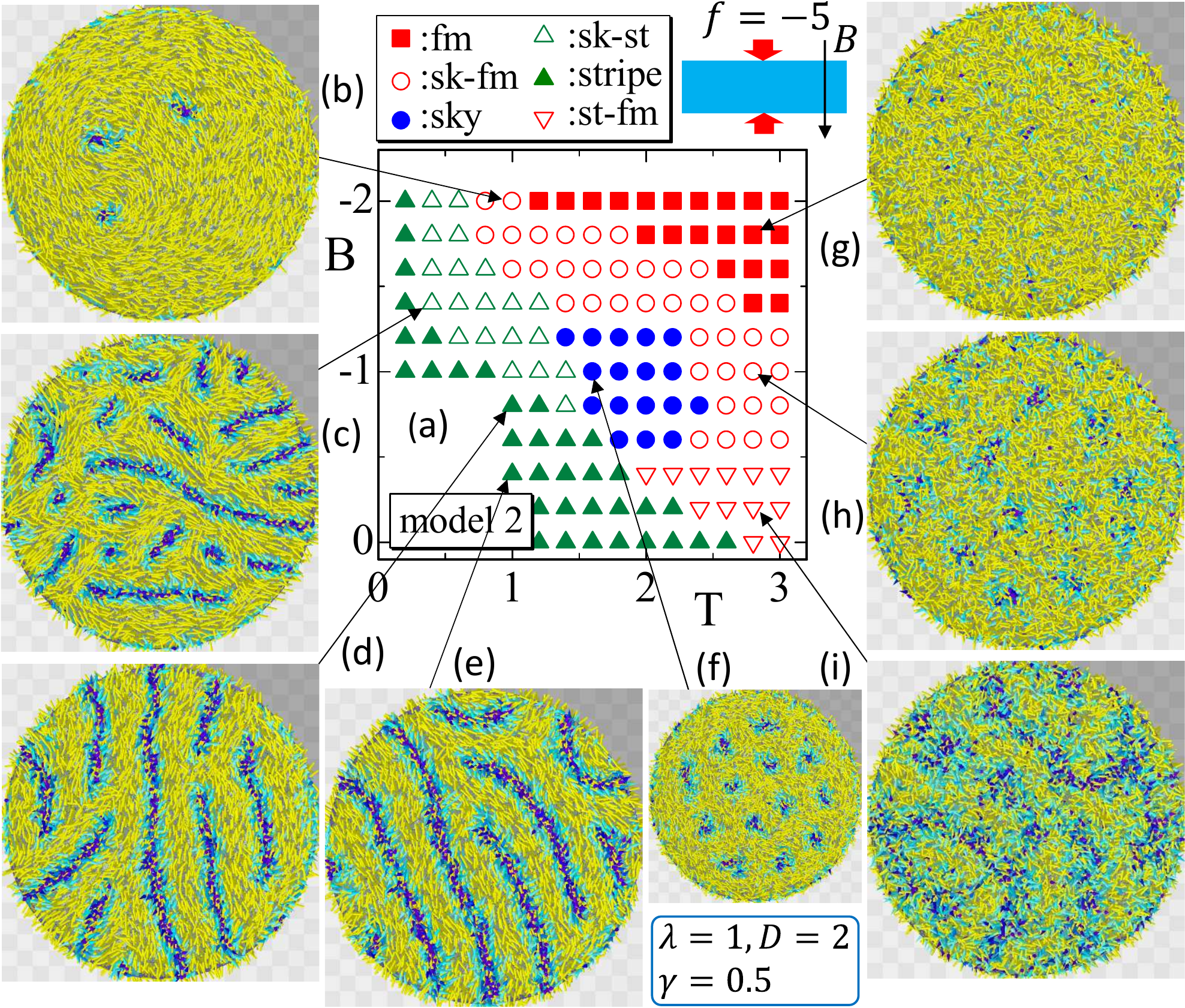} 
\caption {(a)  $B$-$T$ phase diagram and (b)--(i)  snapshots of model 2 under compression  $\vec{f}\!=\!(0,0,-5)$. The other parameters are the same as in Fig. \ref{fig-7}. The area of the skyrmion state  shrinks compared with the case $\vec{f}\!=\!(0,0,0)$ in Fig. \ref{fig-7}(a). Strains $\vec{\tau}$ (\textcolor{Dandelion}{\large{$ -$}}) are perpendicular to $Z$ direction, i.e., in-plane directions, and visible.
\label{fig-9}}
\end{minipage}
\end{tabular}
\end{figure}

Next, we show results obtained under $\vec{f}\!=\!(0,0,-5)$, which corresponds to compression along the  $Z$  direction, as shown by arrows facing each other in Figs. \ref{fig-8} and \ref{fig-9}. 
This situation corresponds to the condition in Fig. \ref{fig-1}(c).
The parameters, except $\vec{f}$, are the same as assumed in Figs. \ref{fig-6} and \ref{fig-7}. We find that the area of skyrmion in the phase diagram shrinks compared with those for $\vec{f}\!=\!(0,0,0)$ in Figs. \ref{fig-6} and \ref{fig-7} in both models. Not only the area of skyrmion but also the total number of skyrmions significantly reduces compared with those in Figs. \ref{fig-6} and \ref{fig-7}.  For this reason, all of the skyrmion states can be identified with the skyrmion-ferromagnetic state, however, we identify the skyrmion state in comparison with other states in the same phase diagram. 
We also find that the strain directions $\vec{\tau}$ denoted by small cylinders (\textcolor{Dandelion}{\large{$ -$}}) are almost vertical to $Z$ direction though its in-plane direction is not always fixed. Nevertheless, it is interesting to see that the stripe directions in Figs. \ref{fig-8}(d),(e) and  \ref{fig-9}(d),(e) align to some specific directions along which $\vec{\tau}$ aligns. These helical directions are considered to appear spontaneously under the random nature of the 3D tetrahedral lattice. 
The shrinkage of the area of the skyrmion state  is consistent with the experimentally confirmed result in Refs.  \cite{Pfleiderer-etal-Science2009,Nii-etal-NatCom2015,Charcon-etal-PRL2015,Seki-etal-PRB2017}. 

\begin{figure}[h]
\begin{tabular}{cc}
\begin{minipage}{0.5\linewidth}
\centering{}\includegraphics[width=8cm]{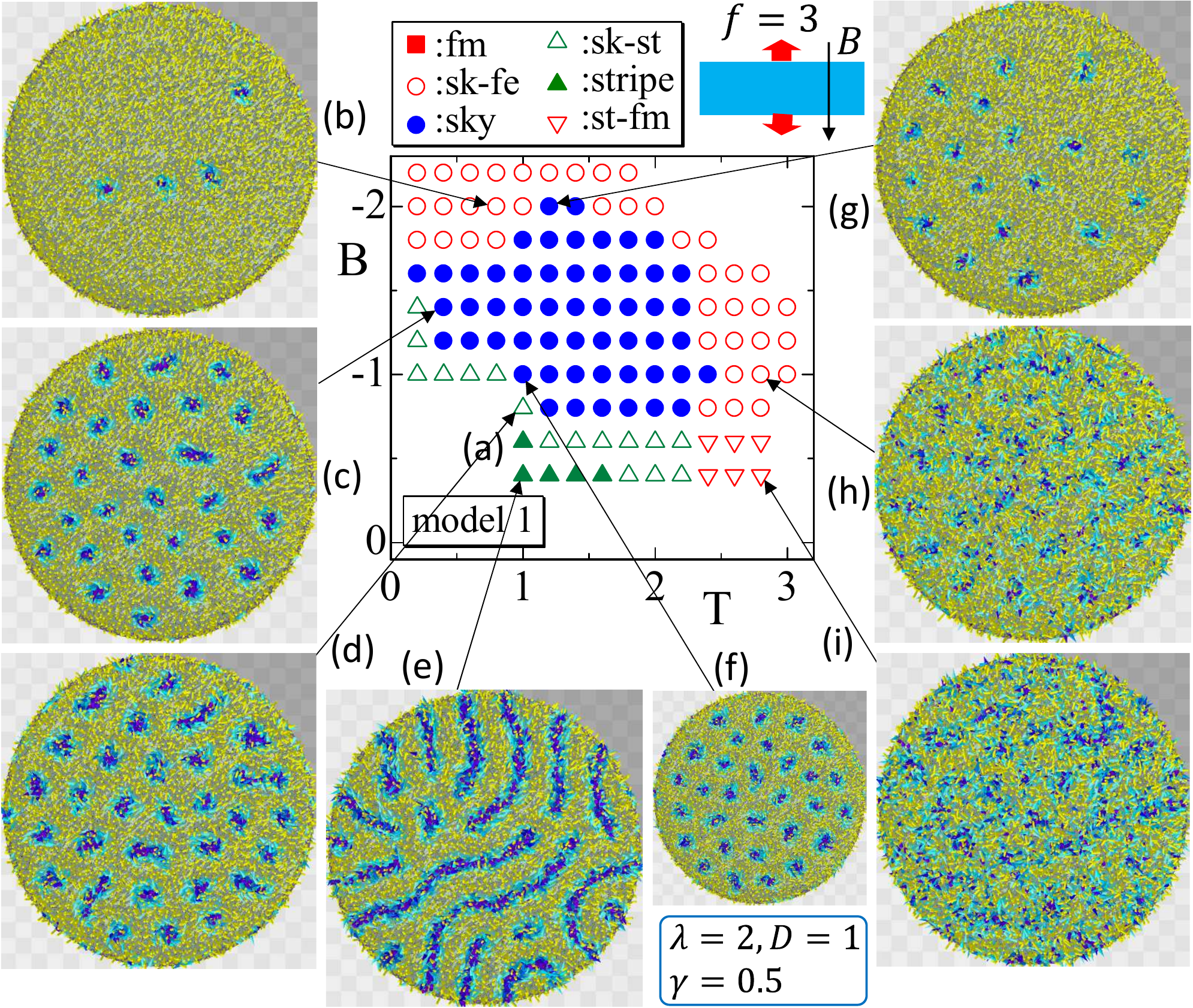}
\caption {(a)  $B$-$T$ phase diagram and (b)--(i)  snapshots of model 1 under $\vec{f}\!=\!(0,0,3)$. The other parameters are the same as in Fig. \ref{fig-8}. The area of the skyrmion state  is comparable to the case $\vec{f}\!=\!(0,0,0)$ in Fig. \ref{fig-6}(a). Strains $\vec{\tau}$ (\textcolor{Dandelion}{\large{$ -$}}) are parallel to $Z$ direction, and almost invisible.
\label{fig-10} }
\end{minipage}
\hspace{0.02\linewidth}
\begin{minipage}{0.5\linewidth}
\centering{}\includegraphics[width=8cm]{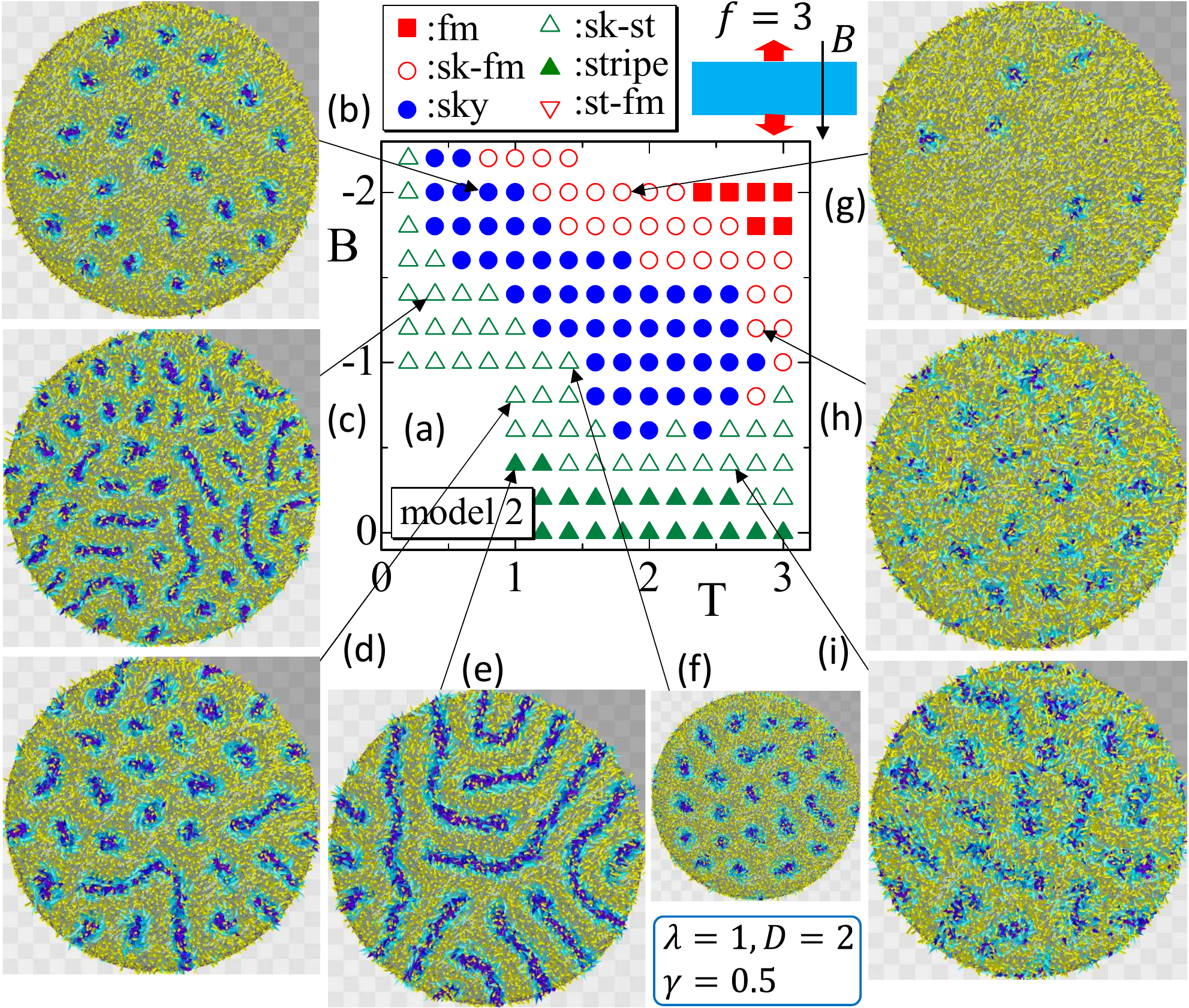} 
\caption {(a)  $B$-$T$ phase diagram and (b)--(i)  snapshots of model 2 under tension $\vec{f}\!=\!(0,0,3)$. The other parameters are the same in Fig. \ref{fig-9}. The area of the skyrmion state  increases compared with the case $\vec{f}\!=\!(0,0,0)$ in Fig. \ref{fig-7}(a).Strains $\vec{\tau}$ (\textcolor{Dandelion}{\large{$ -$}}) are parallel to $Z$ direction, and almost invisible.
\label{fig-11}}
\end{minipage}
\end{tabular}
\end{figure}

Thus, so far no difference is observed between the results of FMI anisotropy in model 1 and DMI anisotropy in model 2. 
Finally, we show the results obtained under  tension $\vec{f}\!=\!(0,0,3)$ in both models in Figs. \ref{fig-10} and \ref{fig-11}. The positive $f(=\!3)$ implies that the stress is tension, as indicated by arrows in the figures.  This situation corresponds to the condition in Fig. \ref{fig-1}(b), where a tensile strain is induced along $Z$ direction. In other words, this tensile stress $f(=\!3)$  along the  $Z$ direction is  expected to induce a tensile strain, which is induced by a compressive stress perpendicular to the $Z$ direction, as those assumed in experiments  \cite{Pfleiderer-etal-Science2009,Nii-etal-NatCom2015,Charcon-etal-PRL2015,Seki-etal-PRB2017}. This point is discussed in Section \ref{Hamiltonian-partition-function}. 
The strain directions $\vec{\tau}$ (\textcolor{Dandelion}{\large{$ -$}})  in the snapshots are almost parallel to $Z$ direction in the case of  $\vec{f}\!=\!(0,0,3)$ in both model 1 and model 2. 
However, we  find that the area of skyrmion of model 1 in Fig. \ref{fig-10} is almost the same as that of the zero stress case in Fig. \ref{fig-6}. 
This result is inconsistent with the experimental results shown in Fig. \ref{fig-1}(b) implying that the magnetoelastic effect is not suitably implemented in model 1. In contrast, the area of skyrmion of model 2 in Fig. \ref{fig-11}(a)  increases compared with that in Fig. \ref{fig-7}(a). 
Thus, 
a clear difference is observed in the results of model 1 and model 2, and  
 the result of model 2 is only consistent with the experimental data obtained for compressions perpendicular to the direction of the magnetic field. 

\subsubsection{Magnetic field vs. Stress diagram of model 2\label{BS-diagram} }
\begin{figure}[h]
\begin{tabular}{cc}
\begin{minipage}{0.5\linewidth}
\centering{}\includegraphics[width=8cm]{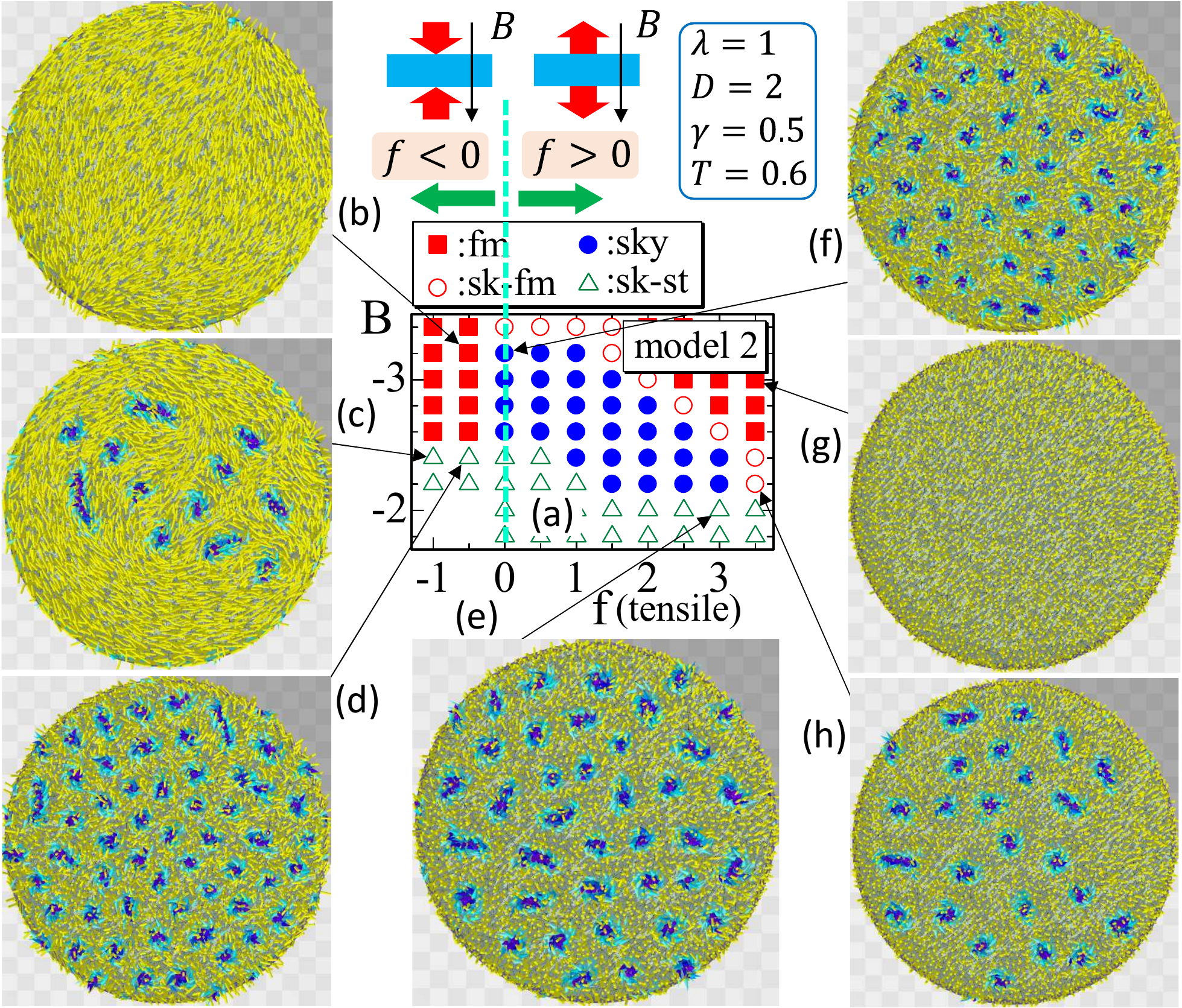} 
\caption {(a) $B$-$f$ phase diagram and (b)--(h)  snapshots of model 2 for tensile stresses ($\Leftrightarrow f>0$), in which the data points for $f\!=\!0$, $f\!=\!-0.5$ and $f\!=\!-1$ are included. Skyrmion-stripe and ferromagnetic phases at $f\!<\!0$ changes to skyrmion phase  for $f\!\geq\!0$ implying that skyrmions are stabilized by tensile stresses, and abrupt changes between ferromagnetic and skyrmion phases can be seen on the vertical dashed line in the region $-3.2\!\leq\! B\!\leq\! -2.6$.
\label{fig-12}}
\end{minipage}
\hspace{0.02\linewidth}
\begin{minipage}{0.5\linewidth}
\centering{}\includegraphics[width=8cm]{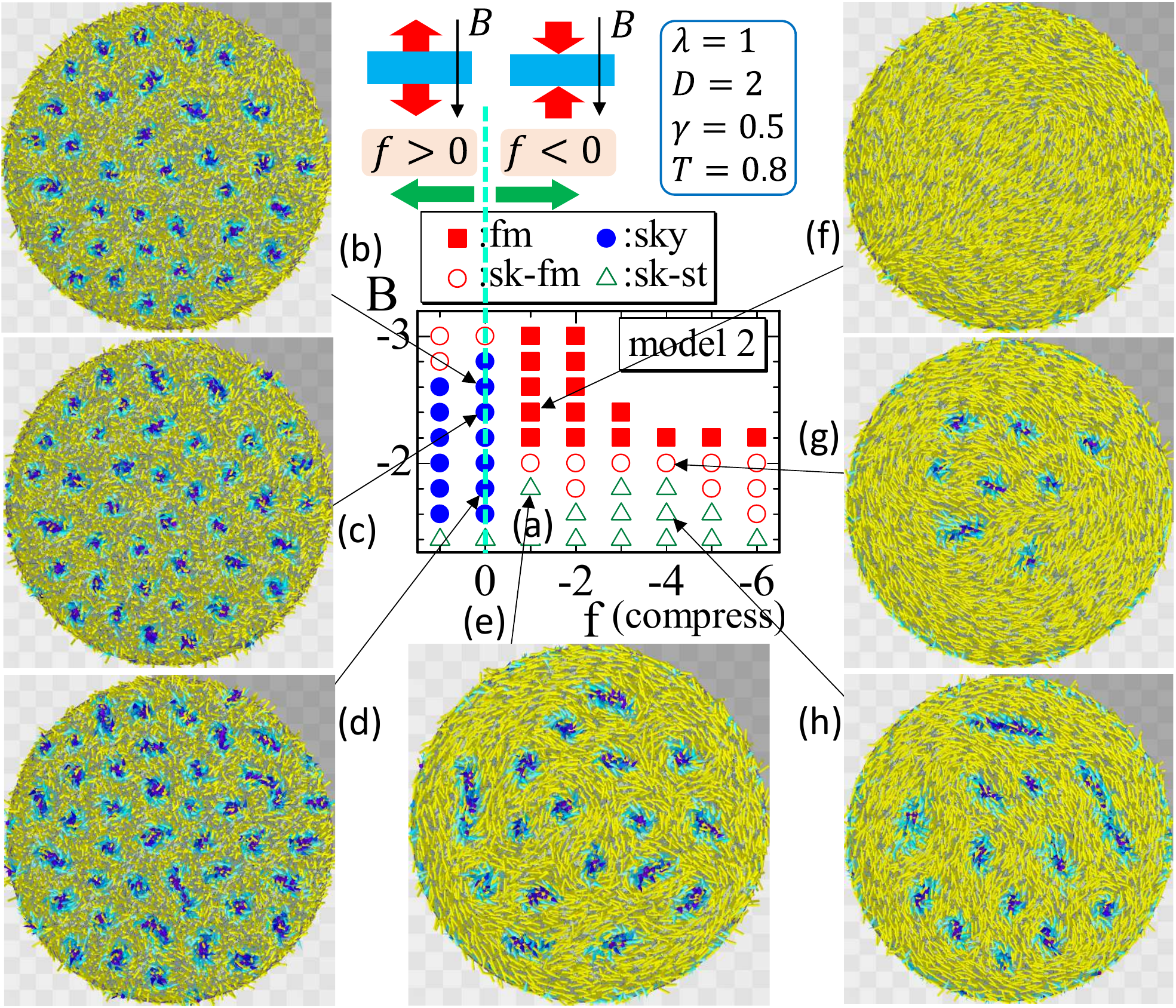}
\caption {(a) $B$-$f$ phase diagram and (b)--(h)  snapshots of model 2 for compressions ($\Leftrightarrow f<0$), in which the data points for $f\!=\!0$ and $f\!=\!1$ are included. Skyrmion phase at $f\!=\!0$ for $|B|\!\geq\! 1.6$ disappear for $f\!\leq\!-1$. Compressive stresses destabilize the skyrmion phase, and abrupt changes can be seen in the region $-2.8\!\leq\! B\!\leq\! -2.2$ between the ferromagnetic and skyrmion phases on the vertical dashed line.
\label{fig-13} }
\end{minipage}
\end{tabular}
\end{figure}
To further confirm that the tensile $f>0$ (compressive $f<0$) stresses stabilize (destabilize) skyrmions, we show $B$-$f$ phase diagrams and snapshots in Figs. \ref{fig-12} and \ref{fig-13} of model 2, in which only DMI is anisotropic \cite{Nii-etal-NatCom2015}. The temperature is fixed to $T\!=\!0.6$ (Fig. \ref{fig-12}) and $T\!=\!0.8$ (Fig. \ref{fig-13}), which are indicated by the dashed lines in Fig. \ref{fig-7}(a) for $f\!=\!0$. Since these $T$ in model 2 are close to a phase boundary of skyrmion region for $f\!=\!0$, we expect skyrmions to appear (disappear) by applying tensile $f\!>\!0$ (compressive $f\!<\!0$) stresses along $Z$ direction, which is  parallel to the direction of $\vec{B}\!=\!(0,0,-B)$. 
The assumed parameters are exactly the same as those in Fig. \ref{fig-7}(a) except $f$, which is varied in the range of  $-1\leq f\leq 3.5$ mainly for tensions (Fig. \ref{fig-12})  and $1\geq f\geq -6$ mainly for compressions (Figs. \ref{fig-13}). 

 By comparing Fig. \ref{fig-12} and Fig. \ref{fig-7}(a), we find  that  the skyrmion-stripe configurations in  Fig. \ref{fig-7}(a) denoted by (\textcolor{green}{$\bigtriangleup$})   on  the dashed line at $T\!=\!0.6$ for   $f\!=\!0$  are stabilized and changed to skyrmions in the region of $1\!<\!f\!\leq3$ and $-2.4\!\leq\!B\!\leq\!-2.2$. 
The total number of skyrmions in  Fig. \ref{fig-12} is clearly increased from those in Fig. \ref{fig-7}(a), as we confirm from the snapshots. 
 The skyrmion phase appears in  Fig. \ref{fig-12}  for $f\!=\!0$ in the range of $-2.6\geq B\geq -3.2$. However, these skyrmions disappear and the ferromagnetic phase appears in the region $B\!\leq\!-2.6$ for negative $f\!\leq\!-0.5$, and, an abrupt change can be seen (Figs. \ref{fig-12}(b) and \ref{fig-12}(f)).  This is understood to be a transition caused by the uniaxial stress $f$ associated with experimentally observed phase transitions  \cite{Pfleiderer-etal-Science2009,Nii-etal-NatCom2015, Charcon-etal-PRL2015,Seki-etal-PRB2017}.   
  To summarize, non-skyrmion configurations  on the dashed line at $T\!=\!0.6$ in Fig. \ref{fig-7}(a)  for $f\!=\!0$ change to skyrmions  if the tensile stresses $f$ are applied by varying from negative ($f\!<\!0$) to positive ($f\!>\!0$) parallel to $\vec{B}\!=\!(0,0,-B)$.

 In the same way, we find from Fig. \ref{fig-13} that the skyrmions on the dashed line at $T\!=\!0.8$ in Fig. \ref{fig-7}(a) in the range of $-1.6\geq B$ for $f\!=\!0$ disappear 
 if $f$ decreases in the negative region. Such an abrupt change, from Fig. \ref{fig-13}(c) to Fig. \ref{fig-13}(f), is expected for a relatively large negative region of $B$ in the range of $-2.8\!\leq\! B\!\leq\! -2.2$ on the dashed line $f\!=\!0$ like in Fig. \ref{fig-12}.  Moreover, a skyrmion configuration in Fig. \ref{fig-13}(d) for $f\!=\!0$ changes to skyrmion-stripe state (Fig. \ref{fig-13}(e)) for  $f\!=\!-1$.  We also find from these two snapshots that the total number of skyrmions significantly changes. This confirms that compressive stresses along $Z$ direction destabilize skyrmion configurations. To summarize, skyrmions on the dashed line at $T\!=\!0.8$ in Fig. \ref{fig-7}(a) almost disappear if compressive stresses are applied parallel to $\vec{B}\!=\!(0,0,-B)$.

\subsubsection{Temperature dependence of physical quantities\label{T-dependence}}
\begin{figure}[h]
\centering{}\includegraphics[width=10.5cm]{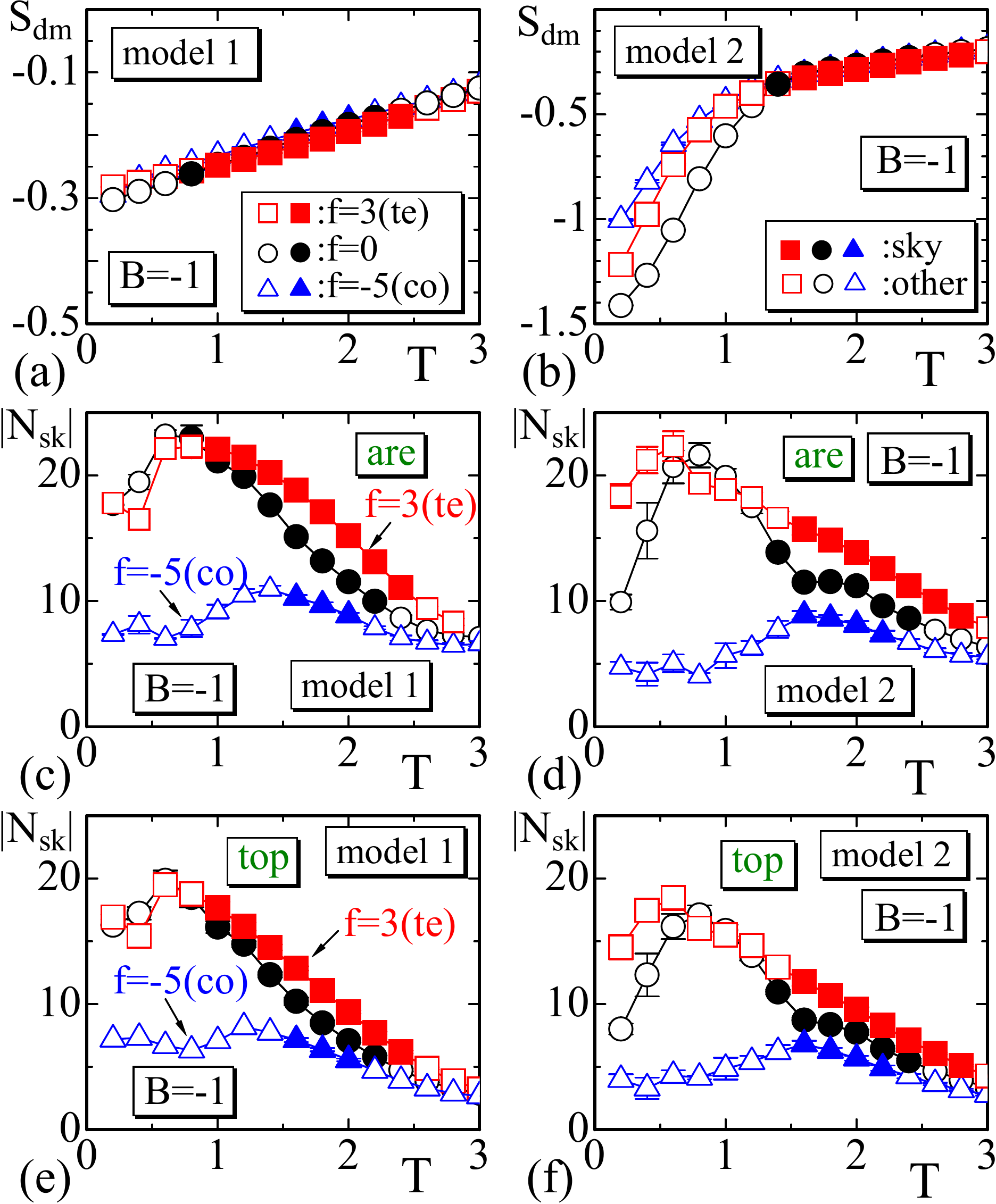}
\caption {DMI energies (a) $S_{\rm dm}\!=\!S_{\rm DM}/N_B$ vs. $T$ of model 1 and (b) $S_{\rm dm}\!=\!S_{\rm DM}/N_{\rm tet}$ vs. $T$ model 2, and the total number $N_{\rm sk}$(are) of skyrmions of (c) model 1 and (d) model 2, and $N_{\rm sk}$(top) of (e) model 1 and (f) model 2, where $N_B$ and $N_{\rm tet}$ are the total number of bonds and tetrahedrons, respectively, and $N_{\rm sk}$ is calculated by Eq. (\ref{skyrmion-number-area}). $S_{\rm dm}$ in (a) is the standard Hamiltonian in Eq. (\ref{model-1}).  The magnetic field is $B\!=\!-1$, and the regions of $T$ indicated by solid (empty) symbols with letters sky (other) are the skyrmion (non-skyrmion) phases.   $N_{\rm sk}$(are) and $N_{\rm sk}$(top) are meaningful  in the skyrmion regions.  (te) and (co) indicate that $f\!=\!3$ and $f\!=\!-5$ correspond to tensile and compressive stresses, respectively.  
\label{fig-14} }  
\end{figure}
In Figs. \ref{fig-14}(a),(b), we plot $S_{\rm dm}\!=\!S_{\rm DM}/N_B$ vs. $T$ of model 1 and $S_{\rm dm}\!=\!S_{\rm DM}/N_{\rm tet}$ vs. $T$ of model 2, where $N_B$ and $N_{\rm tet}$ are the total number of bonds and tetrahedrons, respectively. Note that $S_{\rm dm}$ is negative in both models. This implies that the DMI energy $S_{\rm DM}$ decreases (increases) with increasing (decreasing) DMI coupling coefficient, in contrast to the case of the FMI energy $S_{\rm FM}$, which is always positive. The negative $S_{\rm DM}$ also implies that it makes a counter clock-wise excitation for helical and conical orders along the direction of $\vec{e}_{ij}$ from $\vec{\sigma}_i$ to $\vec{\sigma}_j$ in $S_{\rm DM}$ of Eqs. (\ref{model-1})  and (\ref{model-2})  in the simulations.

The total number $N_{\rm sk}$ of skyrmions is calculated by 
\begin{eqnarray}
N_{\rm sk} ({\rm are})=\frac{1}{4\pi}\sum_{\Delta}a_{\Delta},  \label{skyrmion-number-area}
\end{eqnarray}
where $a_{\Delta}$ is the area on the unit sphere enclosed by three neighboring spins of the triangles on the upper surface of the disk (Fig. \ref{fig-3}(d)). 
This $N_{\rm sk}$ is denoted by ``are''. The number $N_{\rm sk}$ is also calculated by
\begin{eqnarray}
N_{\rm sk}({\rm top})=\frac{1}{4\pi}\int d^2x \vec{\sigma}\cdot \frac{\partial \vec{\sigma}}{\partial x_1} \times \frac{\partial \vec{\sigma}}{\partial x_2}, \label{skyrmion-number-topo}
\end{eqnarray}
which is denoted by ``top''. 
In Figs. \ref{fig-14}(c),(d), and \ref{fig-14}(e),(f), 
we plot $N_{\rm sk}$(are) vs. $T$  and $N_{\rm sk}$(are) vs. $T$ of model 1 and model 2 obtained on the line at $B\!=\!-1$ on the $B$-$T$ diagrams in Figs. \ref{fig-6}(a)-- \ref{fig-11}(a).  Note that $N_{\rm sk}$ is meaningful only in the skyrmion phase. The solid symbols on the figures indicate the range of skyrmions. Thus, we find that  $N_{\rm sk}$ obtained under tensile (compressive) stress is larger (smaller) than that for the zero-stress case in both models. This variation of $N_{\rm sk}$ is consistent with the changes in the area of the skyrmion state in the $B$-$T$ diagrams in Figs. \ref{fig-6}--\ref{fig-11} depending on whether the stress is tensile or compressive at least in model 2; the skyrmion area in the $B$-$T$ diagram is not always increasing for tensile stress in model 1.

Next, we calculate direction dependent coupling constants $\lambda_\mu$  of $S_{\rm FM}$ and $D_\mu$ of $S_{\rm DM}$ from the coefficients $\bar {\Gamma}_{ij}$ by 
\begin{eqnarray}
\begin{split} &  \lambda_{\mu}=(1/N_{\rm tet})\sum_{\it \Delta}\sum_{ij({\it \Delta})}\bar{\Gamma}_{ij}|\vec{e}_{ij}^{\;\mu}|,\quad (\mu=x,y,z),
\end{split}
\label{anisotropy-effective-D}
\end{eqnarray}
where $\sum_{ij({\it \Delta})}$ denotes the sum of six bonds of tetrahedron ${\it \Delta}$ and $N_{\rm tet}\!=\!\sum_{\it \Delta}1$. The $D_\mu$ also has the same expression as $\lambda_\mu$ in Eq. (\ref{anisotropy-effective-D}) though $\bar{\Gamma}_{ij}$ in  $D_\mu$ is different from that in $\lambda_\mu$. >From the definition, $\bar {\Gamma}_{ij}$ is not always identical to $\bar {\Gamma}_{kl}$ for $(i,j)\not=(k,l)$ because $\bar {\Gamma}_{ij}$ has dependence on both the direction from $i$ to $j$ and the position $i$. 
In these two possible dependencies of $\bar {\Gamma}_{ij}$, the direction dependence becomes relevant for anisotropy of interactions if almost all $\tau_i$ align to some specific direction oriented by $\vec{f}$.

\begin{figure}[h!]
\centering{}\includegraphics[width=10.5cm]{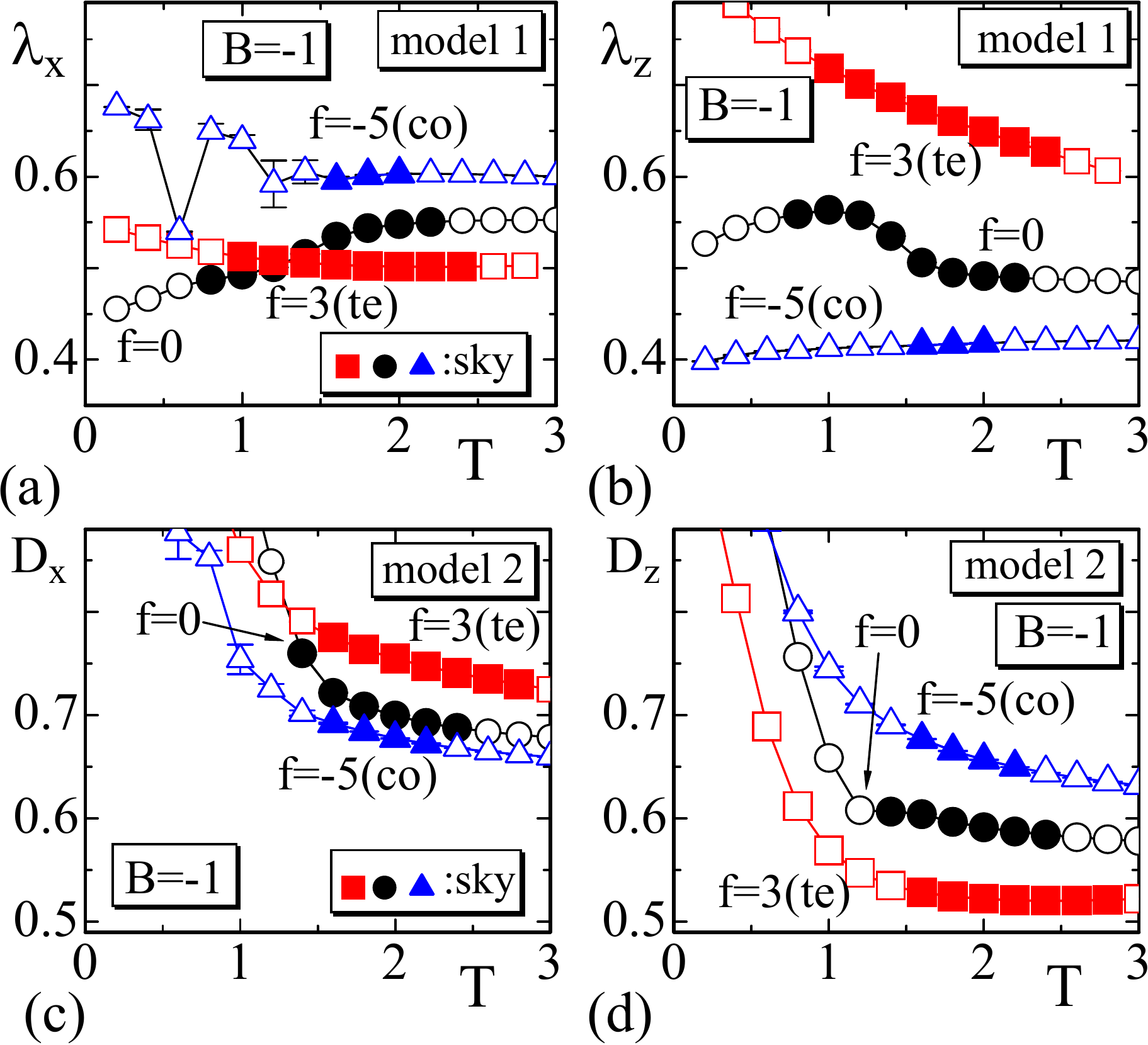}
\caption {Direction dependent coupling constants  in Eq. (\ref {anisotropy-effective-D}) are plotted in (a) $\lambda_x$ and (b) $\lambda_z$ of $S_{\rm FM}$ for model 1, and  in (a) $D_x$ and (b) $D_z$ of $S_{\rm DM}$ for model 2. The regions of $T$ indicated by solid symbols are the skyrmion phases.  The letters (te) and (co) indicate that $f\!=\!3$ and $f\!=\!-5$ correspond to tensile and compressive stresses, respectively. For the tension ($f\!=\!3)$ along $Z$ direction, $\lambda_x$ is small and $\lambda_z$ is large in model 1,  while $D_x$ is large and $D_z$ is small in model 2. These relations between the directions of tensile stress and interaction anisotropy are almost the same as in the 2D models in Ref. \cite{El-Hog-etal-PRB2021}.  The mechanism of the appearance of such anisotropy is briefly discussed in Appendix \ref{E-representation}, showing that DMI anisotropy suitably explains the experimental results in  Refs. \cite{Nii-etal-NatCom2015, Charcon-etal-PRL2015,Seki-etal-PRB2017}. }
\label{fig-15}
\end{figure}
To show that the effective coupling constant becomes direction-dependent, we plot $\lambda_x$ and $\lambda_z$ of model 1 in Figs. \ref{fig-15}(a),(b). We should note that $\lambda_y$ is almost the same as $\lambda_x$, and therefore, only $\lambda_x$ is plotted for simplicity. The symbol ($\bigcirc$) denotes $\lambda_\mu$ corresponding to the zero stress $f\!=\!0$, and (\textcolor{red}{$\square$}) and  (\textcolor{blue}{$\bigtriangleup$}) denote the tension $f\!=\!3$ and compression $f\!=\!-5$, respectively. The $\lambda_x$ is largely fluctuating at $T\!=\! 0.6$ in Fig. \ref{fig-15}(a), however, $(\lambda_x+\lambda_y)/2$  smoothly varies because the fluctuation of $\lambda_y$ is opposite to $\lambda_x$ at this point.  This is why this fluctuation in  $\lambda_x$ does not influence $\lambda_z$ plotted in Fig.\ref{fig-15}(b). The constants $D_x$ and $D_z$ of model 2 are also plotted in Figs.  \ref{fig-15}(c),(d). We find that $\lambda_z(f\!=\!3) >\lambda_z(f\!=\!-5)$ in Fig. \ref{fig-15}(b) along the  $Z$  direction while $\lambda_x(f\!=\!3) < \lambda_x(f\!=\!-5)$  in Fig. \ref{fig-15}(a)  along the $x$ direction in model 1. Such a dependence of effective coupling constant on the sign of $f$ such as  $f>0$ or $f<0$ can also be seen in  $D_\mu$ n Figs. \ref{fig-15}(a),(b) for model 2, and the dependence of  $D_\mu$  is opposite to that of $\lambda_\mu$. From these results, we understand that uniaxial stresses make the interaction strength between spins direction-dependent or anisotropic in both models.  These direction-dependent coupling constants are a non-trivial consequence of the FG modeling, because no magneto-elastic coupling term, such as $\sum_i(\vec{\tau}_i\cdot \vec{\sigma}_i)^2$, is included in the Hamiltonian.

To compare with the results of Ref. \cite{Koretsune-etal-SCRep2015}, in which $D_x$ and $D_y$ are obtained by first-principle calculations, we find that the direction dependence of DMI coefficient in our results is consistent with the reported ones. Indeed, for the tensile stress  $f\!=\!3$(te) along $z$ direction in Figs. \ref{fig-15}(c),(d), we find $D_z < D_x$. This result is consistent with the  reported results that the coefficient $D_y$ along the tensile direction is smaller than $D_x$ in the direction perpendicular to the tensile direction in Ref. \cite{Koretsune-etal-SCRep2015}.

From the dimensional analysis, the ratio $\lambda_x/D_x(\simeq\!\lambda_y/D_x)$ corresponds to the skyrmion size. In model 1, $S_{\rm DM}$ is the standard one and $D_x\!=\!1$, and hence, the size is proportional to $\lambda_x$. Thus, the results $\lambda_x$ in Fig. \ref{fig-15}(a) indicate that the skyrmion size for $f\!=\!3$ (tension) is smaller than that for $f\!=\!-5$ (compression). This implies that the tension $f\!=\!3$ (compression $f\!=\!-5$) stabilizes (destabilizes) skyrmions in model 1. Therefore, this expectation in the skyrmion size for stabilization should be consistent with the presented results in Figs. \ref{fig-6}--\ref{fig-12} for model 1, however this consistency/inconsistency is not always clear because the size also depends on $B$ and the difference in the ratios is not so large. 
On the other hand,  the skyrmion size in model 2 is proportional $1/D_x$ for the same reason, and therefore,  the results $D_x$ in Fig. \ref{fig-15}(c) indicate that the skyrmion size for $f\!=\!3$ (tension) is smaller than that for $f\!=\!-5$ (compression). This expectation should also be seen in the snapshots in Figs. \ref{fig-6}--\ref{fig-11}, however, for the same reason mentioned above, the size difference is not apparent.

We should emphasize that the skyrmion phase is stabilized by the anisotropy in DMI for tensile stress ($f\!=\!3$) applied along $Z$ direction, along which the magnetic field $\vec{B}\!=\!(0,0,-B)$ is applied. 
This relation between the directions of tensile stress and interaction anisotropy is exacltly same in the 2D model of Ref. \cite{El-Hog-etal-PRB2021} except for the fact that $\vec {\tau}$ is prohibited from aligning parallel to $\vec{B}\!=\!(0,0,-B)$ direction in the 2D model because $\vec {\tau}$ has only in-plane components. In the 3D model in this paper, no such limitation is imposed on $\vec {\tau}$, which is a three-component variable. In this sense, the problem of skyrmion stability/instability can only be studied in a 3D model at least in the FG modeling prescription. More detailed information on the reason why and how the strength of interactions depends on the strain direction and the reason why only DMI  in model 2 suitably reproduces results consistent with the experimental data provided in Appendix \ref{E-representation}.

Finally in this subsection, we comment on the numerical results of the standard model with a magneto-elastic coupling defined by $S=\lambda S_{{\rm FM}}+DS_{{\rm DM}}-S_{B}+\gamma S_{\tau}-S_{f}-\alpha \sum_i(\vec{\tau}_i\cdot \vec{\sigma}_i)^2$, in which no FG modeling prescription is assumed, but instead, we add the final term; the magneto-elastic coupling term plays a role in aligning $\vec{\sigma}_i$ along (perpendicular to) $\vec{\tau}_i$ for positive (negative) $\alpha$. As in model 1 and model 2, the direction of $\vec{\tau}$ in this model is controlled by $\vec{f}$. This model also has instability of the skyrmion under compression $f<0$ along the $Z$ direction as in model 1 and model 2. However, this model does not have skyrmion stability. The area of the skyrmion state  in the $B$-$T$ diagram does not increase for positive tensile stresses $f>0$. This inconsistency with the experimental data is the same as in model 1.

\section{Summary and conclusion\label{conclusion} }
We numerically studied skyrmion stability/instability under uni-axial stresses using the Finsler geometry modeling technique.  In this Finsler geometry model, no explicit magneto-elastic coupling term is assumed in the Hamiltonian, while a new degree of freedom $\vec{\tau}$ for strains is introduced as a non-polar three-component variable.  Because of its 3D nature, this variable $\vec{\tau}$ plays a non-trivial role in skyrmion stability/instability, leading to a magneto-elastic effect.

In the simulation study, compression or tensile stress $f$ was applied along the direction perpendicular to  a 3D thin disk, enforcing strains $\vec{\tau}$ to align along or perpendicular to the stress direction. This alignment of $\vec{\tau}$ permits the Dzyaloshinskii-Moriya interaction (DMI) to be dynamically anisotropic. We found that this direction-dependent DMI stabilizes or destabilizes the skyrmion configuration; the area of the skyrmion  state  increases or decreases in the phase diagram of the magnetic field $B$ and the temperature $T$. 

Such an effect of stress $f$ on stability/instability of skyrmions can also be confirmed on the $B$-$f$ phase diagram so that a change between the skyrmion and other non-skyrmion phases depends on whether $f$ is tensile or compressive. These changes in the area of skyrmion state in both $B$-$T$ and $B$-$f$ phase diagrams are consistent with reported experimental results, elucidating that the DMI anisotropy is the origin of the stability/instability.

In the $B$-$f$ phase diagrams, we have demonstrated that the ferromagnetic and skyrmion phases depend separately on the sign of applied mechanical stresses for suitable ranges in temperature and magnetic field. We note that the detailed information provided in Appendix C explains how DMI is modified to be direction-dependent by mechanical stresses, this is helpful to understand the relevant experimental results.

In this paper, the mechanical stresses are implemented in the models by the strain variable $\vec{\tau}$, of which the direction is controlled by external mechanical forces. However, as we demonstrated in Appendix \ref{uniaxial-strains},  the DMI anisotropy is also caused by lattice deformations in the same FG modeling technique, in which the variable $\vec{\tau}$ plays a non-trivial role. For this reason, it is interesting to study the magneto-elastic effect in the same FG model by combining both the external force and the lattice deformation for a modification of $\vec{\tau}$. 

\acknowledgements 
The author H.K. acknowledges Madoka Nakayama and Sohei Tasaki for the helpful discussions. This work is supported in part by a Collaborative Research Project of the Institute of Fluid Science (IFS), Tohoku University, JSPS Grant-in-Aid for Scientific Research 19KK0095, and JSPS Grant-in-Aid for Scientific Research on Innovative Areas "Discrete Geometric Analysis for Materials Design": Grant Number 20H04647. The simulations were partly performed on the computers at the IFS computer center.

\appendix

\section{Construction of the 3D disk \label{3D-disk}}
\begin{figure}[t]
\centering{}\includegraphics[width=11.5cm]{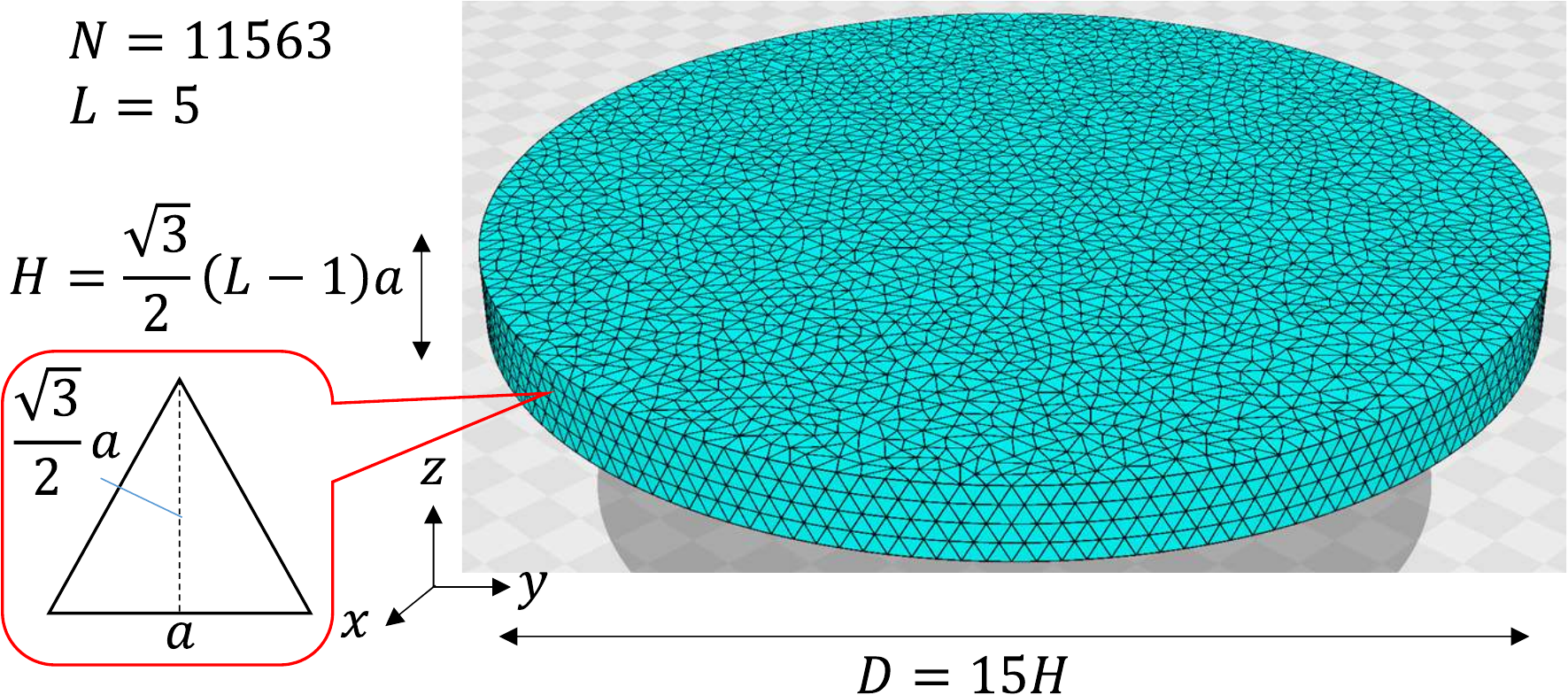}
\caption{A 3D disk of $N\!=\!11563$, which is the total number of vertices. The side wall or surface is composed of a regular triangle of edge length $a$. The height $H$ or thickness of the disk is $H\!=\! \sqrt{3}/2)(L-1)a$, and the diameter is $D\!=\!15H$, where the total number of layers is $L\!=\!5$ on the side surface. The layers or regular triangles appear only on the side surface, where the vertices are regularly distributed.  }
\label{fig-A1} 
\end{figure}

Here, we show detailed information on how to construct a 3D disk, on which a discrete Hamiltonian is defined (Fig. \ref{fig-A1}). The 3D disk is composed of tetrahedrons and reflects no crystalline structure. A crystalline structure is not used because FG modeling has to start with a continuous Hamiltonian, which is described in Appendix \ref{discrete-FMDM}. In the FG modeling technique, local coordinate axes are necessary for the discretization process of a continuous Hamiltonian. The tetrahedron edges play a role in the local coordinate axes. Even if a regular lattice such as a cubic one is used for the 3D volume, the edges  are assumed to be local coordinate axes, although the cubic lattices are expected to play a role in crystalline axes. Therefore, no crystalline structure is introduced in such a discrete model on 3D tetrahedral lattices for effective interactions obtained by discretizing a continuous Hamiltonian.

The first step in the construction is to make a 2D cylindrical lattice of height $H\!=\!(\sqrt{3}/2) (L-1)a$ and diameter $D\!=\!15H$ triangulated by a regular triangle of side length $a$, which is called lattice spacing. We assume $L\!=\!5$ for the disk in Fig. \ref{fig-A1} implying that the thickness of the disk is 4 $\times$ triangle of side length $a$. At this stage, only vertex points are generated on the cylindrical surface. The second step is to distribute vertices randomly on the lower surface of the disk by imposing a minimum distance between two different vertices. The total number of points on this disk surface including the circular edge is determined by $A_{\rm disk}/a_{\rm V}$, where $A_{\rm disk}(=\!\pi D^2)$ is the area of the disk and $a_{\rm V}\!=\!(\sqrt{3}/2)a^2$ is the vertex area given by $(A_{\it \Delta}/3)\!\times\!6$ with the triangle area $A_{\it \Delta}\!=\!(\sqrt{3}/2)a^2$. The generated vertices on the lower disk surface are used also for the upper disk. The third step is to distribute vertices randomly inside the 3D disk enclosed by the side cylinder and the two 2D disks by imposing a minimum distance between two different vertices. The total number of vertices inside is $ n_{\rm tet}V_{\rm disk}/(V_{\rm tet}/3)$, where $n_{\rm tet}$ corresponds to a mean value of the total number of tetrahedrons sharing a vertex, and $V_{\rm disk}$ and $V_{\rm tet}$ are volume of the disk and the regular tetrahedron of side length $a$. The assumed value of $n_{\rm tet}$ is approximately $n_{\rm tet}\!=\!4\pi/\Theta_4(\simeq 22.8)$, where $\Theta_4$ is the solid angle of the regular tetrahedron of edge length $a$. As we see in Fig. \ref{fig-A1}, the vertices are regularly located on the side surface, and all other vertices are distributed uniform-randomly, which implies that these are generated by imposing a minimum distance between randomly distributed vertices in each distribution step.

The final step is to link the vertices by Voronoi tessellation technique \cite{Friedberg-Ren-NPB1984}. The total number of vertices $N$, the total number of bonds $N_B$, the total number of triangles $N_T$ and the total number of tetrahedrons $N_{\rm tet}$ are given by $(N,N_B,N_T,N_{\rm tet})\!=\!(11563,73391,118289,56460)$ for the disk in Fig. \ref{fig-A1}. These numbers satisfy $N\!-\!N_B\!+\!N_T\!-\!N_{\rm tet}\!=\!1$, which is identical to that of a tetrahedron, indicating that all the vertices are linked with bonds to form tetrahedrons. Note that the meshing pattern of the upper disk surface is the same as that of the lower surface because the distribution of the vertices is the same on both disk surfaces, even though these are parts of different configurations of tetrahedrons.

\section{Discretization of ferro-magnetic and Dzyaloshinskii-Moriya interactions in a 3D disk \label{discrete-FMDM}}
\begin{figure}[t]
\centering{}\includegraphics[width=8.5cm]{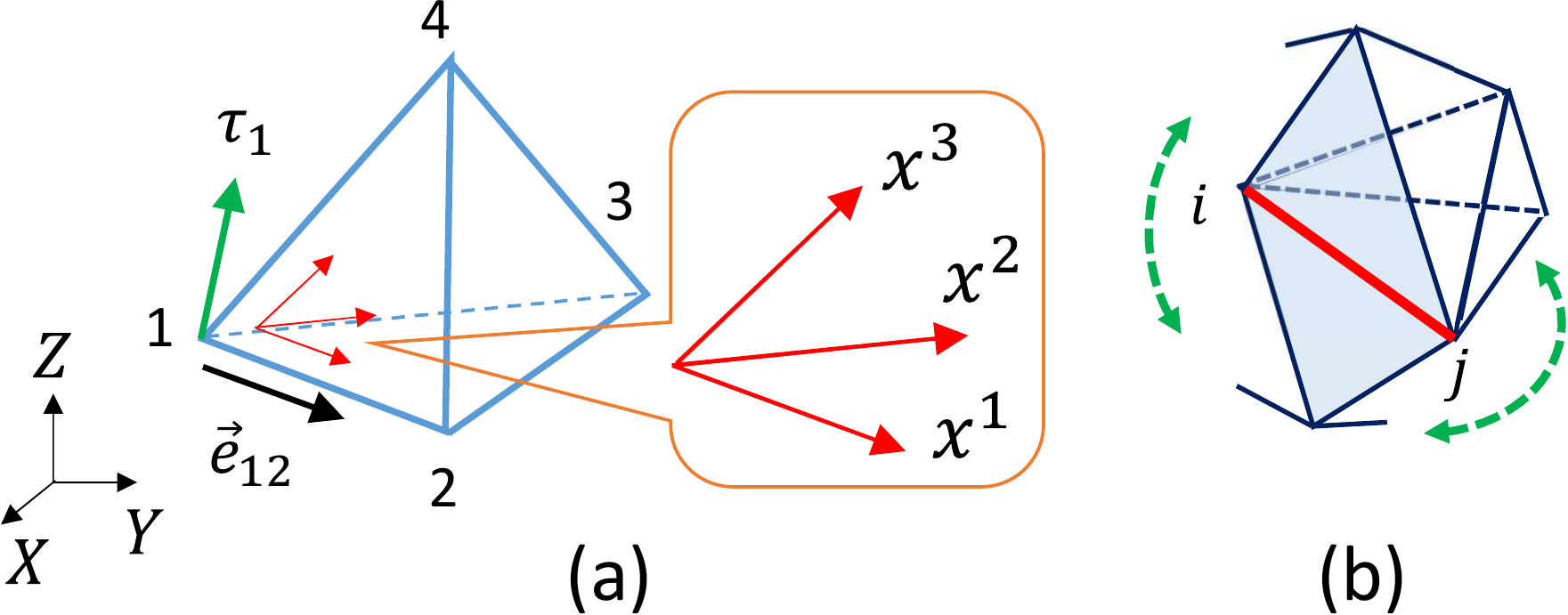}
\caption{(a) A terahedron of vertices 1,2,3,4 with a local coordinate origin at vertex 1, where strain field $\tau_1$ is defined, and $\vec{e}_{12}$ denotes the unit tangential vector from vertex 1 to vertex 2. (b) Tetrahedrons sharing the bond $ij$. The total number of these tetrahedrons is denoted by $n_{ij}(=\!\sum_{{\it \Delta}(ij)}1)$.}
\label{fig-A2} 
\end{figure}

Discrete FMI and DMI energies are obtained from the continuous forms:
\begin{eqnarray}
\label{cont_FDM}
\begin{split} & S_{{\rm FM}}=\frac{1}{2}\int\sqrt{g}d^{3}xg^{ab}\frac{\partial\vec{\sigma}}{\partial x^{a}}\cdot\frac{\partial\vec{\sigma}}{\partial x^{b}},\\
 & S_{{\rm DM}}=\int\sqrt{g}d^{3}xg^{ab}\frac{\partial{\vec{r}}}{\partial x^{a}}\cdot\vec{\sigma}\times\frac{\partial\vec{\sigma}}{\partial x^{b}},
\end{split}
\end{eqnarray}
which are the same as those in a 2D model except for the integration dimension $d^3x$ over local coordinates $x^a(a\!=\!1,2,3)$. 
The metric tensor, the determinant and the inverse are written as
\begin{eqnarray}
\label{F-metric}
g_{ab}=\begin{pmatrix}v_{12}^{-2} & 0 & 0\\
0 & v_{13}^{-2} & 0 \\
0 & 0 & v_{14}^{-2}
\end{pmatrix},
\quad \sqrt{g}=\sqrt{\det g_{ab}}=v_{12}^{-1}v_{13}^{-1}v_{14}^{-1}, \quad g^{ab}=(g_{ab})^{-1},
\label{Finsler-metric}
\end{eqnarray}
where $v_{ij}$ is the unit Finsler length defined by \cite{El-Hog-etal-PRB2021}
\begin{eqnarray}
\label{F-length}
 v_{ij}=\left\{ \begin{array}{@{\,}ll}
                 |\vec{\tau}_{i}\cdot{\vec{e}}_{ij}|+v_{0},&  ({\rm for}\; S_{\rm FM}) \\
                 \sqrt{1-\left(\vec{\tau}_{i}\cdot{\vec{e}}_{ij}\right)^{2}}+v_{0}, &  ({\rm for}\; S_{\rm DM}) 
                  \end{array} 
                   \right.   \\       
 \end{eqnarray}
which are defined on the local coordinate axis $ij$ along the direction from $i$ to $j$ (Fig. \ref{fig-A2}(a)). 
The symbol $\vec{\tau}_i(\in S^2/2: {\rm half\; sphere})$ is the strain field, which is non-polar, and ${\vec{e}}_{ij}$ is the unit tangential vector from vertex $i$ to vertex $j$. The parameter $v_0(>0)$ is introduced as a cutoff for the elements of $g_{ab}$ to be finite. Note that $v_0$ plays a role in the strength of anisotropy. Indeed, for sufficiently large $v_0$, we have $v_{ij}\simeq v_0$, which implies that $g_{ab}$ is constant independent of $\vec{\tau}_i$,  leading to the isotropic interaction. For sufficiently small $v_0$, $v_{ij}$ is dominated by the term $\sqrt{1\!-\!\left(\vec{\tau}_{i}\cdot{\vec{e}}_{ij}\right)^{2}}$ or $|\vec{\tau}_{i}\cdot{\vec{e}}_{ij}|$, which depends considerably on $\vec{\tau}_i$, leading to an anisotropic interaction.

 By replacing the differentials with differences in $S_{\rm FM}$ such that $\partial_1\vec{\sigma}\!\to\! \vec{\sigma}_2\!-\!\vec{\sigma}_1$,  $\partial_2\vec{\sigma}\!\to\! \vec{\sigma}_3\!-\!\vec{\sigma}_1$ and  $\partial_3\vec{\sigma}\!\to\! \vec{\sigma}_4\!-\!\vec{\sigma}_1$ for a local coordinate system with the origin at the vertex $1$, and by replacing $\int\sqrt{g}d^{3}x$ with the sum over tetrahedrons $\sum_{\it \Delta}$ and by using the metric tensor in Eq. (\ref{F-metric}), 
we have $S_{\rm FM}=\sum_{\rm tet}S_{\rm FM}({\rm tet})$ with 
$S_{\rm FM}({\it \Delta})\!=\!(1/2)\sqrt{g}(g^{11}[\vec{\sigma}_2\!-\!\vec{\sigma}_1)^2\!+\!g^{22}(\vec{\sigma}_3\!-\!\vec{\sigma}_1)^2\!+\!g^{33}(\vec{\sigma}_4\!-\!\vec{\sigma}_1)^2]$ on the tetrahedron in Fig. \ref{fig-A2}(a). Usng the expressions of $g$ and $g^{ab}$, we obtain
\begin{eqnarray}
\label{D-SFM-1}
S_{\rm FM}({\it \Delta}) = \frac{v_{12}}{v_{13}v_{14}}(1-\vec{\sigma}_1\cdot\vec{\sigma}_2)+\frac{v_{13}}{v_{12}v_{14}}(1-\vec{\sigma}_1\cdot\vec{\sigma}_3)+\frac{v_{14}}{v_{12}v_{14}}(1-\vec{\sigma}_1\cdot\vec{\sigma}_4).
\end{eqnarray}
Since we have four different local coordinate origins $1$, $2$, $3$ and $4$ on the tetrahedron ${\it \Delta}$ in  Fig. \ref{fig-A2}(a), the expression $S_{\rm FM}({\it \Delta})$ can be replaced by summing over all symmetric expressions obtained by the cyclic replacements $1\!\to\!2, 2\!\to\!3, 3\!\to\!4, 4\!\to\!1$ such that
\begin{eqnarray}
\label{D-SFM-2}
\begin {split}
S_{\rm FM}({\it \Delta}) &=\Gamma_{12}(1-\vec{\sigma}_1\cdot\vec{\sigma}_2)+\Gamma_{13}(1-\vec{\sigma}_1\cdot\vec{\sigma}_3)+\Gamma_{14}(1-\vec{\sigma}_1\cdot\vec{\sigma}_4)\\
&+\Gamma_{23}(1-\vec{\sigma}_2\cdot\vec{\sigma}_3)+\Gamma_{24}(1-\vec{\sigma}_2\cdot\vec{\sigma}_4)+\Gamma_{34}(1-\vec{\sigma}_3\cdot\vec{\sigma}_4),
\end {split}
\end{eqnarray}
with 
\begin{eqnarray}
\begin{split}
\label{D-DFMI-coeff}
& \Gamma_{12}=\frac{v_{12}}{v_{13}v_{14}}+\frac{v_{21}}{v_{23}v_{24}},\quad \Gamma_{13}=\frac{v_{13}}{v_{12}v_{14}}+\frac{v_{31}}{v_{34}v_{32}}, \quad
 \Gamma_{14}=\frac{v_{14}}{v_{12}v_{13}}+\frac{v_{41}}{v_{42}v_{43}}, \\
& \Gamma_{23}=\frac{v_{23}}{v_{21}v_{24}}+\frac{v_{32}}{v_{31}v_{34}}, \quad
  \Gamma_{24}=\frac{v_{24}}{v_{21}v_{23}}+\frac{v_{42}}{v_{41}v_{43}},\quad \Gamma_{34}=\frac{v_{34}}{v_{31}v_{32}}+\frac{v_{43}}{v_{41}v_{42}},
\end{split}
\label{Gamma}
\end{eqnarray}
where the multiplicative factor $1/4$ is removed from $\Gamma_{ij}$.

Here we note that the sum over tetrahedrons in $\sum_{\it \Delta}S_{\rm FM}({\it \Delta})$ can be replaced by the sum over bonds $\sum_{ij}$ such that $\sum_{\it \Delta}S_{\rm FM}({\it \Delta})\!=\!\sum_{ij}S_{\rm FM}({ij})$. In this expression, $S_{\rm FM}({ij})$ is given by $S_{\rm FM}({ij})=\sum_{{\it \Delta}(ij)} S_{\rm FM}({\it \Delta}_{ij})$, where $\sum_{{\it \Delta}(ij)}$ denotes the sum over tetrahedrons sharing the bond $ij$ and $S_{\rm FM}({\it \Delta}_{ij})$ is of the form $S_{\rm FM}({\it \Delta}_{ij})\!=\!\Gamma_{ij}(1\!-\!\vec{\sigma}_i\cdot\vec{\sigma}_j)$ in Eq. (\ref{D-SFM-2}) for the tetrahedron ${\it \Delta}_{ij}$ sharing the bond $ij$ (Fig. \ref{fig-A2}(b)). In this expression $S_{\rm FM}({ij})=\sum_{{\it \Delta}(ij)} S_{\rm FM}({\it \Delta}_{ij})$, it is clear that the interaction between $\vec{\sigma}_i$ and $\vec{\sigma}_j$ depends on the total number of tetrahedrons sharing the bond $ij$, which is given by $n_{ij}\!=\!\sum_{{\it \Delta}_{ij}}1$ even for $\Gamma_{ij}\!=\!1$, implying that 
\begin{eqnarray}
\label{D-SFM-3}
1-\vec{\sigma}_i\cdot\vec{\sigma}_j \not= S_{\rm FM}({ij})=\sum_{{\it \Delta}(ij)} S_{\rm FM}({\it \Delta}_{ij})\quad  {\rm even \;for}\quad \Gamma_{ij}=1(\forall\; ij).
\end{eqnarray}
In fact, we have $S_{\rm FM}({ij})\!=\!n_{ij}(1\!-\!\vec{\sigma}_i\cdot\vec{\sigma}_j )$ on the right hand side for the isotropic case $\Gamma_{ij}\!=\!1$ in the FG modeling prescription.  Note that $\sum_{\it \Delta}\sum_{ij({\it \Delta})}1\!=\!6N_{\rm tet}\!=\!\sum_{ij}\sum_{{\it \Delta}(ij)}1$, where $\sum_{{\it \Delta}(ij)}$ and $\sum_{ij({\it \Delta})}$ denote the sum over tetrahedrons ${\it \Delta}(ij)$ sharing the bond $ij$ and the sum over bonds $ij$ of the tetrahedron ${\it \Delta}$, respectively. 

 We should emphasize that the discrepancy in Eq. (\ref{D-SFM-3}) is acceptable because discrete expressions of Hamiltonian are in general not always uniquely determined. Thus, it is possible for us to use the expression in Eq. (\ref{D-SFM-2}) for $S_{\rm FM}$. Nevertheless, we assume in this paper that the FM interaction is independent of the position in the material whether the position is on the surface or inside, at least for the isotropic case, where $\vec{\tau}_i$ is totally distributed at random.  
 
 For this assumption, we introduce a normalization to $\Gamma_{ij}$ in this paper. The expression of $S_{\rm FM}$ using the normalized factor $\bar {\Gamma}_{ij}$ is given by 
\begin{eqnarray}
\label{D-SFM-4}
\begin{split}
 & S_{\rm FM}=\sum_{{\it \Delta}}\sum_{ij({\it \Delta})}\bar{\Gamma}_{ij}\left(1-\vec{\sigma}_{i}\cdot\vec{\sigma}_{j}\right), \\
 & \bar{\Gamma}_{ij}=\bar{\Gamma}^{-1}\frac{\Gamma_{ij}}{n_{ij}}, \quad \bar{\Gamma}=\frac{\sum_{{\it \Delta}}\left(\sum_{ij({\it \Delta})} \Gamma^0_{ij}/n_{ij}\right)}{\sum_{{\it \Delta}}\sum_{ij({\it \Delta})}1/n_{ij}},
\end{split}
\end{eqnarray}
where  $n_{ij}(=\!\sum_{{\it \Delta}(ij)}1)$ is the total number of tetrahedrons sharing the bond $ij$ as mentioned above. The coefficient $\bar {\Gamma}_{ij}$ is given by dividing ${\Gamma_{ij}}/{n_{ij}}$ by a constant $\bar{\Gamma}$, which is defined by the mean value of $\Gamma^0_{ij}$ with the weight of $1/n_{ij}$. The factor  $1/{n_{ij}}$ in $\Gamma_{ij}/{n_{ij}}$  plays a role in a normalization of ${\Gamma_{ij}}$ because ${\Gamma_{ij}}(1\!-\!\vec{\sigma}_i\cdot\vec{\sigma}_j)$ appears $n_{ij}$ times in $S_{\rm FM}=\sum_{{\it \Delta}}\sum_{ij({\it \Delta})}{\Gamma_{ij}}(1\!-\!\vec{\sigma}_i\cdot\vec{\sigma}_j)$ for each bond $ij$ described as above. The implication of this replacement of $\Gamma_{ij}$ with ${\Gamma}_{ij}/n_{ij}$ is that the interaction of spins between $\vec{\sigma}_i$ and $\vec{\sigma}_j$ becomes independent of $n_{ij}$, which is the total number of spins connected to both  $\vec{\sigma}_i$ and $\vec{\sigma}_j$ (Fig. \ref{fig-A2}(b)). 
 The symbol $\Gamma^0_{ij}$ in the constant $\bar{\Gamma}$ denotes the mean value of  $\Gamma_{ij}$ obtained by 1000 randomly or isotropically distributed configurations of $\vec{\tau}$ with the weight of 1. The value of $\bar{\Gamma}$ for $v_0\!=\!0.1$ is approximately $\bar{\Gamma}\!=\!7.07$ for model 1 and  $\bar{\Gamma}\!=\!2.82$ for model 2. 
 
To explain the definition of $\bar{\Gamma}_{ij}$ and the role of the coefficient $\bar{\Gamma}$ in Eq. (\ref{D-SFM-4}), we consider the case where $v_0$ is sufficiently large such that $v_0\!\gg\!1$. In this case, we have $v_{ij}\!\simeq\!v_0$ from Eq. (\ref{F-length}), and hence, $\Gamma_{ij}$ in Eq. (\ref{D-DFMI-coeff}) is also assumed to be a constant such that $\Gamma_{ij}\!\simeq\! 2/v_0$. Since $\Gamma^0_{ij}$ in $\bar{\Gamma}$ is also  $\Gamma^0_{ij}\!\simeq 2/v_0$, we have   $\bar{\Gamma}\!\simeq\!2/v_0$, and therefore,  $\bar{\Gamma}_{ij}\!=\! \bar{\Gamma}^{-1}{\Gamma_{ij}}/{n_{ij}}\!\simeq\!1/n_{ij}$. From this and Eq. (\ref{D-SFM-4}), we have $S_{\rm FM}\!=\!\sum_{{\it \Delta}}\sum_{ij({\it \Delta})} \bar{\Gamma}_{ij}\left(1\!-\!\vec{\sigma}_{i}\cdot\vec{\sigma}_{j}\right)\!\simeq\!\sum_{{\it \Delta}}\sum_{ij({\it \Delta})} 1/n_{ij}\left(1\!-\!\vec{\sigma}_{i}\cdot\vec{\sigma}_{j}\right)$. Recalling that $\sum_{\it \Delta}\sum_{ij({\it \Delta})}$ denotes the sum over tetrahedrons  ${\it \Delta}$ and the sum over six  bonds $ij({\it \Delta})$ of ${\it \Delta}$, and $n_{ij}\!=\!\sum_{{\it \Delta}(ij)}1$ is the total number of tetrahedrons sharing the bond $ij$, we find that the expression $\sum_{{\it \Delta}}\sum_{ij({\it \Delta})} 1/n_{ij}\left(1\!-\!\vec{\sigma}_{i}\cdot\vec{\sigma}_{j}\right)$ is identical to $\sum_{ij}\left(1\!-\!\vec{\sigma}_{i}\cdot\vec{\sigma}_{j}\right)$, where 
the summation convention is changed from the sum over tetrahedrons ${\it \Delta}$ to the sum over bonds $ij$ by using  the above mentioned relation $\sum_{{\it \Delta}}\sum_{ij({\it \Delta})} \!=\!\sum_{ij} \sum_{{\it \Delta(ij)}}$ and $1/n_{ij}\!=\!1/\sum_{{\it \Delta}(ij)}1$. Since this expression $\sum_{ij}\left(1\!-\!\vec{\sigma}_{i}\cdot\vec{\sigma}_{j}\right)$ is used in the standard definition for FMI, we confirm that the definition of ${\bar \Gamma}_{ij}$ in Eq. (\ref{D-SFM-4}) allows $S_{\rm FM}$ to be identical with the standard expression in the limit of sufficiently large $v_0\!\gg\!1$, in which FMI is isotropic. The problem is how the expression $\bar{\Gamma}_{ij}$ is simplified in the isotropic case that $\vec{\tau}$ is randomly distributed.
For this general case of randomly distributed $\vec{\tau}$, we have $\Gamma_{ij}\!\simeq\!\Gamma_{ij}^0$, and therefore, the average of ${\bar \Gamma}_{ij}$ with the weight $1/n_{ij}$ is given by $\langle {\bar \Gamma}_{ij}\rangle\!=\!\bar{\Gamma}^{-1}\sum_{\it \Delta} \sum_{ij({\it \Delta})}({\Gamma}_{ij}^0/n_{ij})/\sum_{\it \Delta} \sum_{ij({\it \Delta})}(1/n_{ij})\!=\!\bar{\Gamma}^{-1}\bar{\Gamma}\!=\!1$ by the definition of ${\bar \Gamma}_{ij}$ in Eq. (\ref{D-SFM-4}).

 The constant $\bar{\Gamma}$ in $\bar{\Gamma}_{ij}$ can be replaced with $\tilde{\Gamma}\!=\!{\sum_{{\it \Delta}}\left(\sum_{ij({\it \Delta})} \Gamma^0_{ij}/n_{ij}\right)}/{\sum_{{\it \Delta}}\sum_{ij({\it \Delta})}1}$, which is the mean value of $\Gamma^0_{ij}/n_{ij}$ with the weight of 1. In this case, the mean value of $\tilde{\Gamma}^{-1}\Gamma_{ij}/n_{ij}$, which can be written as $\tilde{\Gamma}_{ij}\!=\!\tilde{\Gamma}^{-1}\Gamma_{ij}/n_{ij}$, satisfies $\langle \tilde{\Gamma}_{ij} \rangle\!=\!1$ for $\Gamma_{ij}\!=\!\Gamma_{ij}^0$,  where the symbol $\langle Q \rangle$ for physical quantity $Q$ is defined by $\langle Q \rangle \!=\!{ \sum_{\it \Delta} \sum_{ij({\it \Delta})} Q} / {\sum_{\it \Delta} \sum_{ij({\it \Delta})} 1}$. Indeed, we have $\langle \tilde{\Gamma}^0_{ij} \rangle\!=\!\langle\tilde{\Gamma}^{-1}\Gamma^0_{ij}/n_{ij} \rangle\!=\!\tilde{\Gamma}^{-1}{ \sum_{\it \Delta} \sum_{ij({\it \Delta})} ({\Gamma}^0_{ij}/n_{ij}}) /{\sum_{\it \Delta}\sum_{ij({\it \Delta})}  1}\!=\!1$. We note that the property $\langle \tilde {\Gamma}^0_{ij}\rangle\!=\!1$ is independent of  $v_0$ as mentioned above, where  $v_0$ corresponds to the interaction anisotropy,  though $\Gamma_{ij}$ in Eq. (\ref{Gamma}), and hence, $\bar \Gamma_{ij}$ and  $\tilde \Gamma_{ij}$ depend on $v_0$. Note also that the difference between  $\bar{\Gamma}$ and $\tilde{\Gamma}$ is a constant factor, which can be absorbed in the coefficient $\lambda$ of $S_{\rm FM}$. For this reason, no difference is expected in the final results obtained by the two different coupling constants $\bar \Gamma_{ij}$ and  $\tilde \Gamma_{ij}$.

The discrete expression for $S_{\rm DM}$ is given by
\begin{eqnarray}
\label{D-SDM-1}
 S_{\rm DM}=\sum_{{\it \Delta}}\sum_{ij({\it \Delta})} \bar{\Gamma}_{ij}{\vec{e}}_{ij}\cdot\vec{\sigma}_{i}\times\vec{\sigma}_{j},
\end{eqnarray}
which is obtained by the same procedure as for $S_{\rm FM}$. The expression of $\bar{\Gamma}_{ij}$ in $S_{\rm DM}$ is the same as that in $S_{\rm FM}$ in Eq. (\ref{D-SFM-4}), however, the values are different from each other due to the difference in $v_{ij}$, as shown in Eq. (\ref{F-length}).

\section{Finsler geometry modeling of skyrmions\label{E-representation}}
\subsection{Mathematical descriptions \label{M-framework}}
\begin{figure}[th]
\centering{}\includegraphics[width=13.5cm]{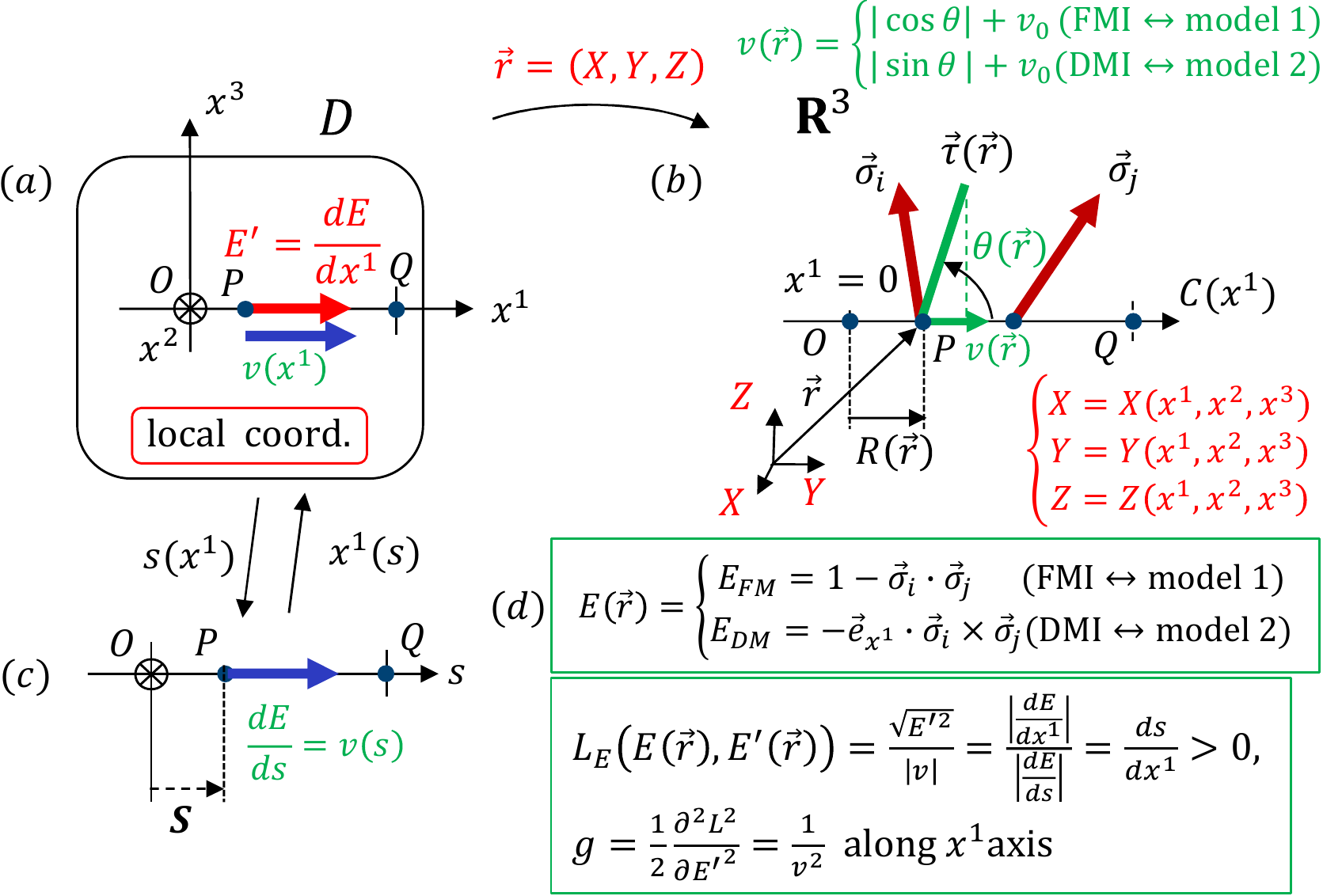}
\caption{Illustrations of a mapping $\vec{r}(X,Y,Z)$ a mathematical description of FG modeling of FMI in model 1 and DMI in model 2. (a) A local coordinate axis $x^1$ at $O$ in a three-dimensional Euclidean domain $D$, of which the $x^1$ axis corresponds to (b) the line ${\mathcal C}(x^1)$ a part of material  ($\subset\! {\bf R}^3$) by the mapping $\vec{r}$, where spins $\vec{\sigma}_i$ and $\vec{\sigma}_j$ are located at the positions $\vec{r}(P)$ and its neighborhood, and $\vec{\tau}(\in S^2/2: {\rm half\;sphere})$ denotes a tensile stress direction. (c) A line ($\subset\! D$) parametrized by $s$, which are connected with $x^1$ axis by coordinate transformations $s(x^1)$ and $x^1(s)$ each other. (d) $E(\vec{r})$ is the discrete FMI or DMI energy along the line from $P$ to the neighboring point in Eq. (\ref{Energies-FMI-DMI}), and Finsler function $L_E\!=\!|E^\prime|/|v|$ and the corresponding Finsler metric $g\!=\!v^{-2}$ along $x^1$ axis. The positions of $\vec{\sigma}_i$ and $\vec{\sigma}_j$ correspond to vertices 1 and 2 on the $x^1$ axis of the tetrahedron in Fig. \ref{fig-A2}(a).
\label{fig-A3}}
\end{figure}

DMI anisotropy is reported to be the origin of skyrmion shape-deformation and stability/instability \cite{Shibata-etal-Natnanotech2015,Koretsune-etal-SCRep2015,Seki-etal-PRB2017}. Here we show how the effective FMI and DMI coefficients of models 1 and 2 in Eq. (\ref{anisotropy-effective-D}) depend on the tensile strain, or in other words, how magnetoelastic effects are implemented in FMI of model 1 and DMI of model 2.  The problems are; how do mechanical strains cause an interaction anisotropy in both model 1 and model 2? Why does only the DMI anisotropy in model 2 suitably explain the experimental results?
 To consider these problems, we try to intuitively illustrate the technical details of the discrete FG modeling of FMI and DMI in Figs. \ref{fig-A3}(a)--(d) and briefly describe it first. 
 
  As shown in Fig. \ref{fig-A2}(a), materials are divided into tetrahedrons, where a local coordinate is introduced for a discretization of Hamiltonians.  In Fig.  \ref{fig-A3}(a), we show a local coordinate with the origin $O$. Here we consider a Finsler function only along the $x^1$ axis for simplicity.  In Fig. \ref{fig-A3}(b), a straight line ${\mathcal C}(x^1)$ is drawn inside the material. This line corresponds to the $x^1$ axis of the tetrahedron in Fig. \ref{fig-A2}(a) and includes the point $\vec{r}(P)$ and its neighborhood, at which  spins  $\vec{\sigma}_i, \vec{\sigma}_j (\in S^2: {\rm unit\; sphere})$ and a tensile strain $\vec{\tau}(\in S^2/2: {\rm half\;sphere})$ are defined.  Let $\vec{e}_{x^1}$ be the unit tangential vector  corresponding to $\vec{e}_{12}$ in Fig. \ref{fig-A2}(a). Using this $\vec{e}_{x^1}$, we define  $\vec{v}(\vec{r})\!=\!v(\vec{r})\vec{e}_{x^1}$  with ${v}(\vec{r})\!=\!|\cos\theta|\!+\!v_0$ (FMI in model 1) and ${v}(\vec{r})\!=\!|\sin\theta|\!+\!v_0$ (DMI in model 2) to implement interaction anisotropy, where $\theta$ is the angle between $\vec{\tau}$ and $\vec{e}_{x^1}$, and $v_0(=\!0.1)$ is a small number (see Eq. (\ref{F-length})). Along the line ${\mathcal C}(x^1)$, we define energy functions $E(\vec{r})$ at $\vec{r}$ by the line integral of interactions along ${\mathcal C(x^1)}$ from ${\vec r}(O)$ to ${\vec r}(P)$ such that 
 \begin{eqnarray}
  \label{Cont-Energies-FMI-DMI}
E(\vec{r})=\left\{ \begin{array}{@{\,}ll}
                 E_{\rm FM}=\frac{1}{2}\int_{0}^{x^1}(\frac{\partial \vec{\sigma}}{\partial x^1})^2 dx^1 (>0)& \; ({\rm FMI}  \; \leftrightarrow\;{\rm model \; 1}) \\
                 E_{\rm DM}=-\int_{0}^{x^1}\frac{\partial \vec{r}}{\partial x^1}\cdot \vec{\sigma}\times\frac{\partial \vec{\sigma}}{\partial x^1} dx^1 (>0)& \;  ({\rm DMI}\;  \leftrightarrow\;{\rm model \; 2}) 
                  \end{array} 
                   \right.,      
\end{eqnarray}
where $x^1(O)$ and $x^1(P)$ are written as $x^1(O)\!=\!0$ and $x^1(P)\!=\!x^1$ for simplicity. Note that the inequality $(>\!0)$ can be replaced by $(\geq\! 0)$  in Eq. (\ref{Cont-Energies-FMI-DMI}). 
 These expressions can also be written by using the discrete form  of FMI or DMI   such that 
 \begin{eqnarray}
\label{Energies-FMI-DMI}
E(\vec{r})=\left\{ \begin{array}{@{\,}ll}
                 E_{\rm FM}=\sum_{(ij)\in {\mathcal C}}\left(1-\vec{\sigma}_i\cdot \vec{\sigma}_j\right)(>0)& \; ({\rm FMI}  \; \leftrightarrow\;{\rm model \; 1}) \\
                 E_{\rm DM}=-\sum_{(ij)\in {\mathcal C}}\left(\vec{e}_{x^1}\cdot\vec{\sigma}_i\times \vec{\sigma}_j\right) (>0)& \;  ({\rm DMI}\;  \leftrightarrow\;{\rm model \; 2}) 
                  \end{array} 
                   \right.,       
\end{eqnarray}
which is monotonically increasing with respect to $x^1$ and can be used to define a new local coordinate or parametrization.  
Note that, from the expressions of  $E(\vec{r})$ in Eqs. (\ref{Cont-Energies-FMI-DMI}) and (\ref{Energies-FMI-DMI}),  the derivatives $E^\prime(=\!dE/dx^1)$ of $E(\vec{r})$ along $x^1$ direction are given by 
 $E^\prime_{\rm FM}\!=\!(1/2)({\partial \vec{\sigma}}/{\partial x^1})^2\!=\!1-\vec{\sigma}_i\cdot \vec{\sigma}_j$ and $E^\prime_{\rm DM}\!=\!-\vec{e}_{x^1}\cdot\vec{\sigma}_i\times \vec{\sigma}_j$, where $i$ and $j$ are nearest neighbor sites on the $x^1$ axis. The reason why such an energy $E(\vec{r})$ along the axis $x^1$ is considered meaningful is that anisotropy of interaction can be reflected in $E(\vec{r})$ if the interaction is direction dependent.

  Let $s$ be another parametrization of ${\mathcal C}(x^1)$ satisfying
\begin{eqnarray}
\label{E-derivative}
\frac{dE}{ds}=v(s)=\left\{ \begin{array}{@{\,}ll}
                 |\cos\theta|+v_0& \; ({\rm FMI}  \; \leftrightarrow\;{\rm model \; 1}) \\
                 |\sin\theta|+v_0 & \;  ({\rm DMI}\;  \leftrightarrow\;{\rm model \; 2}) 
                  \end{array} 
                   \right.        
\end{eqnarray}
 (Fig. \ref{fig-A3}(c)), which defines a ``Finsler length'' $s$ along $C(x^1)$, where $E\!=\!E_{\rm FM}$ or $E\!=\!E_{\rm DM}$. The Finsler length $S$ from  $\vec{r}(O)$ to $\vec{r}(P)$ can also be obtained  by the Finsler function
\begin{eqnarray}
\label{F-function-E}
L_E=\frac{\sqrt{(dE/dx^1)^2}}{|v|}=\frac{|dE/dx^1|}{|dE/ds|}=\frac{ds}{dx^1}>0\quad ({\rm on}\; x^1\; {\rm axis})
\end{eqnarray}
(Fig. \ref{fig-A3}(d))  with the Finsler geometry prescription \cite{Matsumoto-SKB1975,Koibuchi-Sekino-PhysA2014,Bao-Chern-Shen-GTM200} such that $S\!=\!\int_{x^1(O)}^{x^1(P)}L_E dx^1, (x^1(O)\!=\!0)$. Note that the differential and integral should be replaced by the difference and discrete sum because of the discrete nature of the modeling, however, we use both of the continuous and discrete expressions for simplicity.  

 This Finsler length $S$ is not actually used in the modeling, but we should note that $S$ is different from the ordinary Euclidean length $R(\vec{r})$ between $\vec{r}(O)$ and $\vec{r}(P)$ (Fig. \ref{fig-A3}(b)). However, if this new parameter is used as a local coordinate for continuous Hamiltonians such as those in Eq. (\ref{cont_FDM}), the new Hamiltonian is expected to have an anisotropy as an effect of another interaction such as magnetoelastic interaction implemented via $v(s)$. For this purpose of the implementation of another interaction, a metric function is sufficient, and we the have Finsler metric along $x^1$ axis from $L_E\!=\!|E^\prime|/|v|$ such that $g\!=\!(1/2)\partial^2 L_E^2/\partial E^{\prime 2}\!=\!1/v^2$  \cite{Koibuchi-Sekino-PhysA2014,Matsumoto-SKB1975,Bao-Chern-Shen-GTM200}.  This $g\!=\!v^{-2}$ corresponds to $v_{12}^{-2}$ in Eq. (\ref{F-metric}). The same procedure can be applied to the other axes $x^2$ and $x^3$ in Fig. \ref{fig-A3}(a) to obtain the Finsler metric in Eq. (\ref{F-metric}).

\begin{figure}[th]
\centering{}\includegraphics[width=10.5cm]{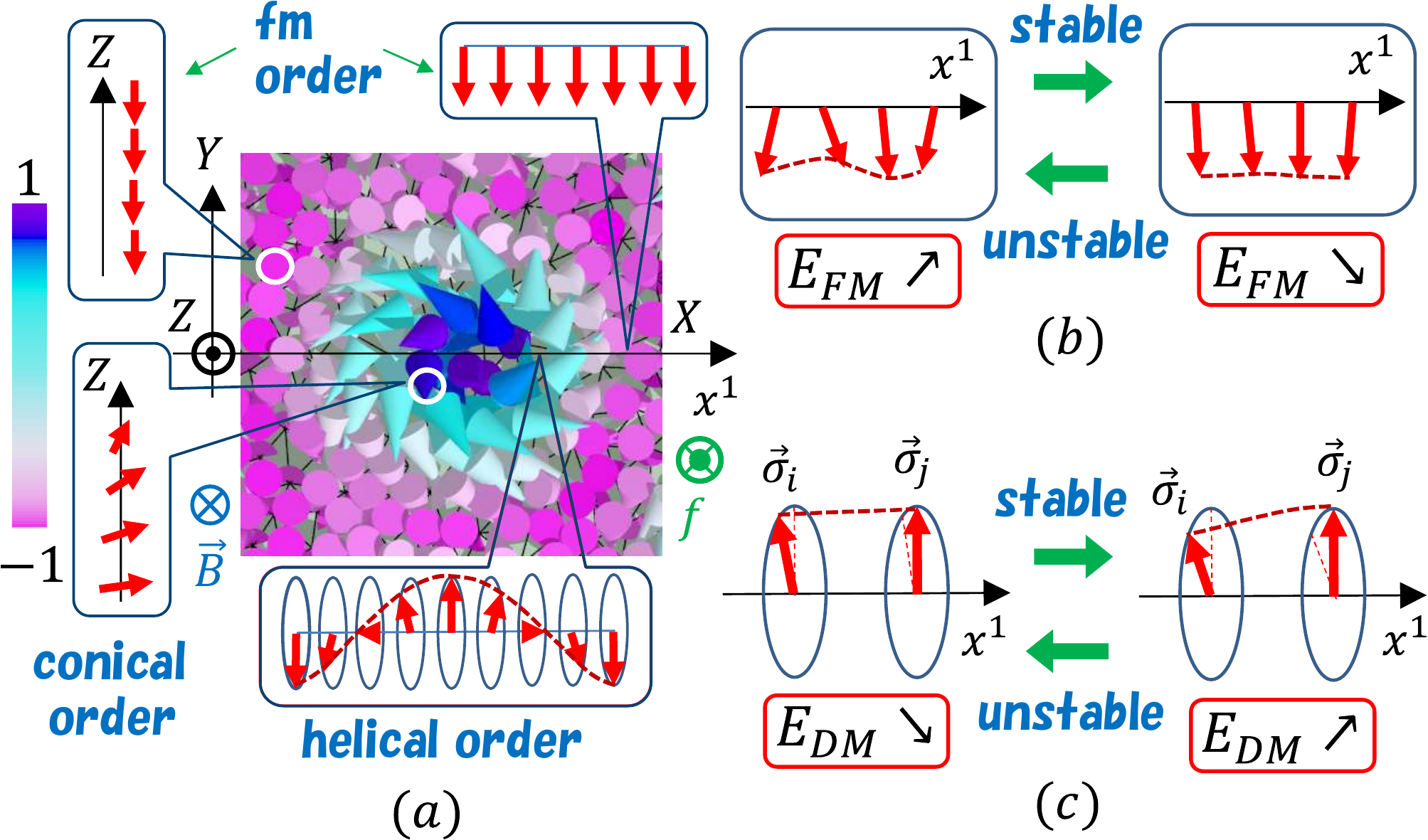}
\caption{(a) A stable skyrmion configuration (= a vacuum state) obtained on the 3D disk,  and an illustration of  left-handed helical, ferromagnetic (fm) and right-handed conical orders along $X$ or $x^1$ axis and $Z$ axis, where a magnetic field $\vec{B}\!=\!(0,0,-B)$ and a tensile ($f\!=\!3$) or compressive ($f\!=\!-5$) stress are applied along $Z$ axis. The illustration of conical order corresponds to those close to the center of skyrmion.  (b) Two different configurations of ferromagnetic orders on  the $x^1$ axis; the configuration on the left  (right) side makes an unstable (stable) contribution to the ferromegnetic order in the skyrmion configuration,  and (c) two different configurations of nearest neighbor spins $\sigma_i$ and $\sigma_j$ in the helical order on the $x^1$ axis; the configuration on the left (right) side makes an unstable (stable) contribution to the helical order in the skyrmion configuration.  Note that  $E_{\rm DM}>0 \Leftrightarrow S_{\rm DM}<0$ because of the definition of $E_{\rm DM}$ in Eq. (\ref{Energies-FMI-DMI}), and $E_{\rm DM}\nearrow $ ($\Leftrightarrow$ DMI stable) means $E_{\rm FM}\nearrow $ ($\Leftrightarrow$ FMI unstable) implying the competition between DMI and FMI. See text for an explanation of the origin of the  ``stable'' and ``unstable''  contributions used in (b) and (c). 
\label{fig-A4}}
\end{figure}

\subsection{Stability and instability of skyrmion configuration \label{SaI-skyrmion}}
 Now, let us consider a skymion configuration under the condition $\theta\simeq \pi/2$ corresponding to the case of a tensile stress applied to the direction of $\vec{B}\!=\!(0,0,-B)$, or equivalently, the case of a compressive stress along $x^1$ axis, which is perpendicular to $\vec{B}\!=\!(0,0,-B)$  (see Fig.\ref{fig-1}(b)).  As it turns out, the skyrmion configuration is stable as shown in Fig. \ref{fig-A4}(a), 
where one observes the stability of the helical order and the ferromagnetic order along $x^1$ axis, and the ferromagnetic and conical orders along $Z$ axis direction. We should note that the helical order is left-handed because $S_{\rm DM}$ is negative in our models 1 and 2 (see Figs. \ref{fig-14}(a),(b)).
Due to the circular symmetry, this stable skyrmion configuration in Fig. \ref{fig-A4}(a) implicitly implies that the $x^1$ axis can be chosen in another in-plane direction. Taking into account the condition $dE/ds\!=\!v$ in Eq. (\ref{E-derivative}),  the situation for a tensile strain $\vec{\tau}$ parallel to $\vec{B}$; i.e., $\vec{\tau}\parallel \vec{B}$, corresponds to $f\!=\!3$ in our simulations. This case is equivalent to the case where the  stress is perpendicular to $\vec{B}$ (see Fig.\ref{fig-1}(b)). Thus, the condition can be summarized as follows:
\begin{eqnarray}
\label{v-pi-half}
\begin{split}
\theta\simeq \frac{\pi}{2} &\Leftrightarrow \frac{dE}{ds} =v \simeq
\left\{ \begin{array}{@{\,}ll}
                 v_0& \; ({\rm FMI} \; \leftrightarrow\;{\rm model \; 1}) \\
                 1+v_0 & \;  ({\rm DMI} \;  \leftrightarrow\;{\rm model \; 2})
                  \end{array} 
                   \right.
\\
& \Leftrightarrow 
 \left\{ \begin{array}{@{\,}ll}
                E_{\rm FM}(=\int v ds)\;\;{\rm is\; small\;along\;}C \;&\;({\rm FMI}\; \leftrightarrow\;{\rm model \; 1}) \\
                E_{\rm DM}(=\int v ds)\;\;{\rm is\; large\;along\;}C \;& \;  ({\rm DMI}\;\leftrightarrow\;{\rm model \; 2}) 
                  \end{array} 
                   \right.,
\end{split}
\end{eqnarray}
where $E_{\rm FM}(=\int v ds\!=\!\int (dE_{\rm FM}/ds) ds\!=\!\int dE_{\rm FM}) ({\rm FMI})$ and $E_{\rm DM}(=\int v ds\!=\!\int (dE_{\rm DM}/ds) ds\!=\!\int dE_{\rm DM}) ({\rm DMI})$ correspond to $1\!-\!\vec{\sigma}_i\cdot\vec{\sigma}_j$ and $-\vec{e_{x^1}}\cdot\vec{\sigma}_i\times\vec{\sigma}_j(=\!|\vec{e_{x^1}}\cdot\vec{\sigma}_i\times\vec{\sigma}_j|)$, respectively,  in the $x^1$ axis between vertices 1 and 2 in Fig. \ref{fig-A2}(a) in our discrete model as shown in Eq. (\ref{Energies-FMI-DMI}).

The statement 
 ``$E_{\rm FM}$ is small along $C$ $({\rm FMI})$'' in Eq. (\ref{v-pi-half})  means  the ferromagnetic orders  along $x^1$ axis outside the skyrmion configuration in Fig. \ref{fig-A4}(a) are mainly stabilized like a configuration change in Fig. \ref{fig-A4}(b). This contributes to stabilization of the skyrmions.  However, ``$E_{\rm FM}$ is small along $C$ $({\rm FMI})$'' in Eq. (\ref{v-pi-half}) also means ``$E_{\rm FM}$ is large along $Z$ axis'' because $v\!\simeq\!|\cos 0|\!=\!1$ (FMI) along $Z$ axis. This implies that the ferromagnetic order along $Z$  destabilizes the skyrmions. Thus, FMI is not satisfactorily anisotropic for stabilizing skyrmions in Fig. \ref{fig-A4}(a) at least in model 1.  

In contrast, the statement ``$E_{\rm DM}$ is large along $C$ $({\rm DMI})$'' in Eq. (\ref{v-pi-half}) means that the DMI energy $E_{\rm DM}$ becomes large, or in other words,  the helical order becomes more stable like a configuration change in Fig. \ref{fig-A4}(c), implying  that the skyrmion size becomes smaller so that the total number of skyrmions is increased.  Note that this remark also explains the reason why the effective  DMI coupling $D_x$ becomes large along $x^1$ axis in the numerical result shown in Fig. \ref{fig-15}(c) for $f\!=\!3$. Indeed, the value of $S_{\rm DM}$ is negative in our modeling (Figs. \ref{fig-14}(a),(b)), and therefore, the helical order is changed to be stable like a change in Fig. \ref{fig-A4}(c) for a larger DMI coupling $D_x$ in a reasonable range. Moreover, the DMI coupling $D_z$ along $Z$ axis is also expected to be small, because ``$E_{\rm DM}$ is large along $C$ $({\rm DMI})$'' implies ``$E_{\rm DM}$ is small along $Z$ axis''. This expectation explains the numerical result that $D_z (f\!=\!3)$ is small in Fig. \ref{fig-15}(d).  This small $D_z$ along $Z$ direction is expected to make  stable the skyrmion in Fig. \ref{fig-A4}(a). In fact, two different conical orders are expected in  3D skyrmions along $Z$ direction; one is close to the center like in Fig. \ref{fig-A4}(a) and the other is slightly far from the center. Since these directions, left-handed or right-handed, are opposite to each other, a large $D_z$ can destroy one of them and therefore has a negative effect on the stable skyrmion. Moreover, the ferromagnetic order, which can be a conical order, in $Z$ direction can also be influenced by such a large $D_z$. 

Furhter information on the stability/instability of skyrmions can be obtained by considering the condition $\theta\!\simeq\!0$, where the direction of tensile stress is along $x^1$ axis, which corresponds to a compressive stress applied along $Z$ direction or parallel to $\vec{B}\!=\!(0,0,-B)$ (see Fig. \ref{fig-1}(c)). Note that the symbol $f$ in Fig. \ref{fig-1}(c) denotes a compressive stress, and the same symbol $f$ is used here for both tensile $f\!>\!0$ and compressive $f\!<\!0$ stresses along $Z$ direction. 
One more point to note is that the compression $f\!<\!0$ does not always imply that $\vec{\tau}$ is parallel to $x^1$ axis but $\vec{\tau}$ can be parallel to $x^2$ axis. Therefore,  ``$x^1$ axis'' can be replaced by ``$x^2$ axis'' in the following discussion  if it is necessary. 
 Then, we have
\begin{eqnarray}
\label{v-zero}
\begin{split}
\theta\simeq 0 &\Leftrightarrow \frac{dE}{ds} =v \simeq
\left\{ \begin{array}{@{\,}ll}
                 1+v_0& \; ({\rm FMI} \; \leftrightarrow\;{\rm model \; 1})\\
                 v_0 & \;  ({\rm DMI} \;  \leftrightarrow\;{\rm model \; 2})
                  \end{array} 
                   \right.
\\
& \Leftrightarrow 
 \left\{ \begin{array}{@{\,}ll}
                E_{\rm FM}(=\int v ds)\;\;{\rm is\; large\;along\;}C \;&\;({\rm FMI}\; \leftrightarrow\;{\rm model \; 1}) \\
                E_{\rm DM}(=\int v ds)\;\;{\rm is\; small\;along\;}C \;& \;  ({\rm DMI}\;\leftrightarrow\;{\rm model \; 2}) 
                  \end{array} 
                   \right.,
\end{split}
\end{eqnarray}
where ``$E_{\rm FM}$ is large along $C$ $({\rm FMI})$''  implies that ferromagnetic orders outside the skyrmion are destabilized along $x^1$ axis. This may be the reason for a small skyrmion area  in the $B$-$T$ phase diagram in Fig. \ref{fig-8}(a) for $f\!=\!-5$ (compared with  Fig. \ref{fig-6}(a) for $f\!=\!0$). However, this ``$E_{\rm FM}$ is large along $C$ $({\rm FMI})$''  also means ``$E_{\rm FM}$ is small along $Z$ axis'' implying that the ferromagnetic order along $Z$ direction is stabilized. Thus, for the same reason as for the condition in Eq. (\ref{v-pi-half}), FMI in model 1 stabilizes ferromagnetic orders only in one direction along the direction of $\vec{\tau}$. This implies that FMI does not properly implement the magnetoelastic effect  for skyrmions, at least from the viewpoint of FG modeling.

In contrast, ``$E_{\rm DM}$ is small along $C$ $({\rm DMI})$'' implies that the helical order becomes unstable (Fig. \ref{fig-A4}(c)), and hence, the skyrmion configuration is destabilized.  This result is consistent with the numerical result shown in Fig. \ref{fig-9}(a) for model 2 and also the experimental fact that compressive stresses parallel to $B$ make skyrmions unstable \cite{Nii-etal-NatCom2015, Charcon-etal-PRL2015,Seki-etal-PRB2017}. 
Moreover, the fact that $E_{\rm DM}$ is small along $x^1$ axis means that the corresponding DMI coupling $D_x$ is small compared with the case of Eq. (\ref{v-pi-half}) for the same reason in the case of large $E_{\rm DM}$ as described above, and this expectation is consistent with the numerical result $D_x$ in Fig. \ref{fig-15}(c) for $f\!=\!-5$. In addition, the fact that $E_{\rm DM}$ is small along $x^1$ axis also implies that $D_z$ becomes large along $Z$ axis because of the same argument made for small $D_z$ in the case of Eq. (\ref{v-pi-half}) in model 2, and this expectation of large $D_z$ for $\theta\!\simeq\!0$ explains the simulation data in Fig. \ref{fig-15}(d) for $f\!=\!-5$.

To conclude, qualitative arguments  corresponding to the two specific conditions in Eqs. (\ref{v-pi-half}) and (\ref{v-zero}) for the skyrmion configuration along $x^1$ axis support that only DMI anisotropy is consistent with the experimental results.

As a last remark, we should note that the Finsler function $L_E(E,E^\prime)$ in Eq. (\ref{F-function-E})  is more suitable than $L(R,R^\prime)\!=\!\frac{\sqrt{(dR/dx^1)^2}}{|v|}=\frac{|dR/dx^1|}{|dR/ds|}=\frac{ds}{dx^1}>0$ with $R(\vec{r})$ (Fig. \ref{fig-A3}(b)) to obtain the Finsler metric for studying  interaction anisotropy, even though this representation $L(R,R^\prime)$ looks more straightforward because $L(R,R^\prime)$ can be introduced to replace the Euclidean distance $R$ with Finsler length $S$. 
This is because $L(E,E^\prime)$ is defined by the ratio of energy change $dE/dx^1$ along $C(x^1)$ axis and $dE/ds$ along  the new coordinate to be defined, where $dE/ds$ reflects another interaction to be involved such as the magnetoelastic interaction. As a consequence, the new parameter along $C(x^1)$ axis is defined such that it is easier to get information on how large/small the energy changes are along the axis, as mentioned in the preceding subsection. Thus, we understand that $L(E,E^\prime)$ is more suitable to see anisotropy of energies, namely the direction dependence of energies such as FMI and DMI, as a response to external stimulus such as mechanical stress. For this reason,  we can use $L_E$ and the corresponding metric together with $E$ and $E^\prime$ to extract  information of directions along which the corresponding energy FMI or DMI anisotropically changes.  Even in the 2D models in \cite{El-Hog-etal-PRB2021}, the expressions of interaction coefficient $\Gamma_{ij}$ are relatively complex, and therefore, its anisotropy can only be extracted numerically from $\Gamma_{ij}$ although the FG modeling technique allows us to consider the effective interaction for each direction.

\section{Responses to uniaxial strains\label{uniaxial-strains} }
In this Appendix, we show that only model 2 is consistent with experimental data in the stripe phase for the response to mechanical stresses obtained  at room temperature with zero magnetic field \cite{JDho-etal-APL2003}. 
We should emphasize here that the role of real strains is in aligning the strain variable $\vec{\tau}$. In both model 1 and model 2, the direction of $\vec{\tau}$ is controlled by the stress $\vec{f}$. Another possibility for aligning $\vec{\tau}$ is a lattice deformation, because $S_{\rm DM}$ in Eqs. (\ref{model-1}) and (\ref{model-2}) depends on lattice shape, which can change the unit tangential vector $\vec{e}_{ij}$ from $\vec{\sigma}_i$ to $\vec{\sigma}_j$. $S_{\rm FM}$ in model 1 is also expected to indirectly depend on lattice deformations, because the coefficient $\bar{\Gamma}_{ij}$ of $S_{\rm FM}$ depends on $\vec{\tau}$ and $S_{\rm DM}$ of model 1 depends on the lattice deformation for the same reason as above. 
Therefore, the directions of $\vec{\tau}$ caused by lattice deformations should be consistent with the real strain direction. In the 2D model in Ref. \cite{El-Hog-etal-PRB2021}, where $\vec{\tau}$ is a two-components in-plane variable, this consistency on the direction of $\vec{\tau}$ holds only in model 2. Thus,  we check whether the direction of $\vec{\tau}$ is consistent with the lattice deformation in 3D model 1 and model 2.

We should note that real strains can be assumed in the 3D disk by using $\xi (\simeq 1)$ such that
\begin{eqnarray}
\label{strains}
(\vec{e}_x,\vec{e}_y,\vec{e}_z)\to (\xi\vec{e}_x,\frac{\vec{e}_y}{\sqrt{\xi}},\frac{\vec{e}_z}{\sqrt{\xi}}),\; {\rm or}\; (\frac{\vec{e}_x}{\sqrt{\xi}},\xi\vec{e}_y,\frac{\vec{e}_z}{\sqrt{\xi}})
\end{eqnarray}
where $\xi\vec{e}_x$ denotes a deformation along $x$ direction by $\xi$. Thus, the volume remains unchanged. We have checked in the case of $\xi\!=\!1.07$ that the stripe direction is consistent (inconsistent) with the experimentally observed direction in model 2 (model 1) like in the 2D models in Ref. \cite{El-Hog-etal-PRB2021}.

This generalized vector ${\vec{e}}_{ij}^{\;\prime}$ is identical
to the original unit vector ${\vec{e}}_{ij}$ for $\xi\!=\!1$.
 Note also that ${\vec{e}}_{ij}$ in $v_{ij}$ in Eqs. (\ref{model-1})
and (\ref{model-2}) is  replaced by ${\vec{e}}_{ij}^{\;\prime}$ as follows:

\begin{eqnarray}
\begin{split} & S_{{\rm FM}}=\sum_{\it \Delta}\sum_{ij({\it \Delta})}\bar{\Gamma}_{ij}\left(1-\vec{\sigma}_{i}\cdot\vec{\sigma}_{j}\right),\quad 
S_{{\rm DM}}=\sum_{ij}{\vec{e}}_{ij}^{\;\prime}\cdot\vec{\sigma}_{i}\times\vec{\sigma}_{j},  \\
 &  v_{ij}=\left\{ \begin{array}{@{\,}ll}
                 |\vec{\tau}_{i}\cdot{\vec{e}}_{ij}^{\;\prime}|+v_{0} &  (|\vec{\tau}_{i}\cdot{\vec{e}}_{ij}^{\;\prime}|<1) \\
                 1+v_0 &  (|\vec{\tau}_{i}\cdot{\vec{e}}_{ij}^{\;\prime}|\geq 1) 
                  \end{array} 
                   \right., \quad({\rm model\;1}), 
\end{split}
\label{model-1-prime}
\end{eqnarray}
and 
\begin{eqnarray}
\begin{split} & S_{{\rm FM}}=\sum_{ij}\left(1-\vec{\sigma}_{i}\cdot\vec{\sigma}_{j}\right),\quad
 S_{{\rm DM}}=\sum_{\it \Delta}\sum_{ij({\it \Delta})}\bar{\Gamma}_{ij}{\vec{e}}_{ij}^{\;\prime}\cdot\vec{\sigma}_{i}\times\vec{\sigma}_{j}, \\
 & v_{ij}=\left\{ \begin{array}{@{\,}ll}
                 \sqrt{1-\left(\vec{\tau}_{i}\cdot{\vec{e}}_{ij}^{\;\prime}\right)^{2}}+v_{0} &  (|\vec{\tau}_{i}\cdot{\vec{e}}_{ij}^{\;\prime}|<1) \\
                 v_0 &  (|\vec{\tau}_{i}\cdot{\vec{e}}_{ij}^{\;\prime}|\geq 1) 
                  \end{array} 
                   \right., \quad({\rm model\;2}),  
\end{split}
\label{model-2-prime}
\end{eqnarray}
where $\Gamma_{ij}$ in Eq. (\ref{Gamma}) is defined by $v_{ij}$ in Eqs. (\ref{model-1-prime}) and  (\ref{model-2-prime}). These definitions of $v_{ij}$ are the same as those in the 2D models in Ref. \cite{El-Hog-etal-PRB2021}.

\begin{figure}[h!]
\centering{}\includegraphics[width=9.5cm]{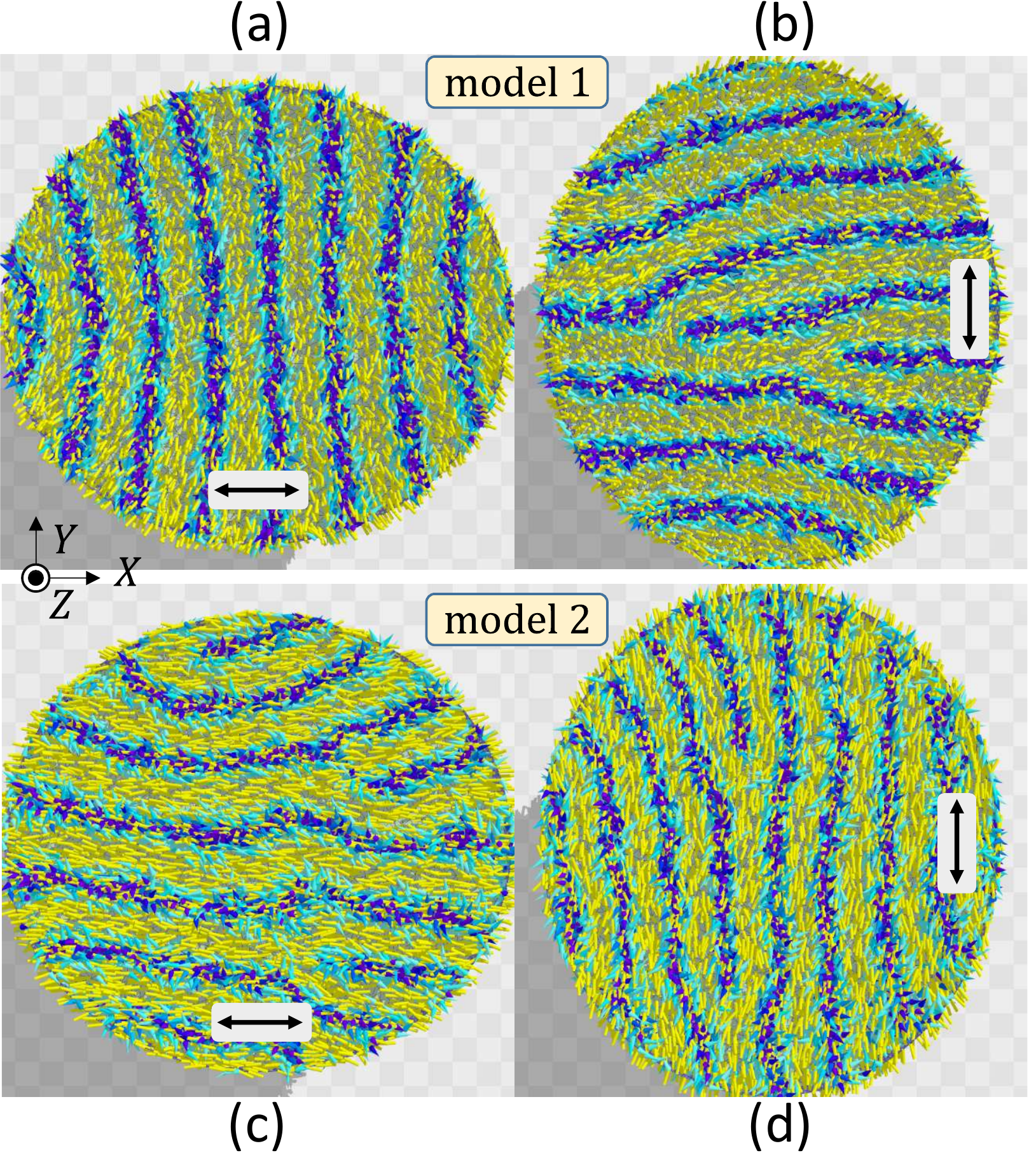}
\caption{Snapshots obtained by (a),(b) model 1 with $(T,\lambda,D,B,\gamma,f)\!=\!(1,2,1,0,1,0)$ and (c),(d) model 2 with $(T,\lambda,D,B,\gamma,f)\!=\!(1,1,2,0,1,0)$.  $v_0\!=\!0.1$, which is the same as in the simulations in Section \ref{results}, is assumed in both models. The shape is given by $\xi\!=\!1.07$  for (a),(c) with $(\xi\vec{e}_x,\frac{\vec{e}_y}{\sqrt{\xi}},\frac{\vec{e}_z}{\sqrt{\xi}})$  and also for (b),(d) with $(\frac{\vec{e}_x}{\sqrt{\xi}},\xi\vec{e}_y,\frac{\vec{e}_z}{\sqrt{\xi}})$. Small cylinders (\textcolor{Dandelion}{\large{$-$}}) denote the directions of $\vec{\tau}$, and the arrows ($\leftrightarrow$, $\updownarrow$) denote the stretched direction. 
The results of model 1 (model 2) are inconsistent (consistent) with the experimental data in \cite{JDho-etal-APL2003}. 
\label{fig-A5}}
\end{figure}

Snapshots obtained by model 1 and model 2 are shown in Figs. \ref{fig-A5}(a)--(d).  The strains of the 3D disk are given by $\xi\!=\!1.07$ in $(\xi\vec{e}_x,\frac{\vec{e}_y}{\sqrt{\xi}},\frac{\vec{e}_z}{\sqrt{\xi}})$ for (a),(c)  and in $(\frac{\vec{e}_x}{\sqrt{\xi}},\xi\vec{e}_y,\frac{\vec{e}_z}{\sqrt{\xi}})$ for (b),(d). The assumed parameters are provided in the figure caption. We should note that the internal strain field $\tau$ is controlled by the deformation of the disk, and therefore, $f$ is fixed to $f\!=\!0$ in this case. 
We find from the snapshots in Figs. \ref{fig-A5}(a) and (b) for model 1 that the direction of $\vec{\tau}$ is perpendicular to the real strain direction;  $\vec{\tau}$ is almost or many of $\vec{\tau}$ are parallel to $Y$ ($X$) axis in Fig. \ref{fig-A5}(a) (Fig. \ref{fig-A5}(b)). In contrast, from the snapshots in Figs. \ref{fig-A5}(c) and (d) for model 2 we find that  the direction of $\vec{\tau}$ is almost parallel to the real strain direction, which is the stretched direction of the lattice.  Thus, we find only in model 2 that the directions of  $\vec{\tau}$ caused by the stretching deformations are consistent with those expected by the tensile stresses $\vec {f}$ applied to the same directions of lattice deformation.  

Finally, we show that only the response of model 2, the direction of helical orders as a response to $\vec{\tau}$ caused by the real strains, is suitably explained by using the notion of $E_{\rm FM}$ and $E_{\rm DM}$  in Appendix \ref{E-representation}. Here we simply use the information on the direction of $\vec{\tau}$ shown in Figs. \ref{fig-A5}(a)--(d). 
The response in model 2 is clear from the stability of the helical orders along the stretched directions indicated by the arrows ($\leftrightarrow, \updownarrow$). The $\vec{\tau}$ direction in Figs. \ref{fig-A5}(c) and (d) are parallel to the arrows, and hence, the helical orders along the directions vertical to the arrows are stabilized for the same reason for the case of skyrmion in  Fig. \ref{fig-A4}(a), where the helical order along $x^1$ axis is stabilized if $\vec{\tau}$ is along $Z$ direction, perpendicular to $x^1$ axis.  To the contrary, in the case of model 1 shown in Figs. \ref{fig-A5}(a) and (b), the $\vec{\tau}$ directions are almost perpendicular to the arrows ($\leftrightarrow, \updownarrow$) and remain in-plane,  and therefore, ferromagnetic orders are stabilized, or in other words $S_{\rm FM}$ becomes small along the arrow directions. This implies a destabilization of helical order along the arrow direction because FMI and DMI should be competitive; one is weak (strong) and the other is strong (weak) for a given configuration such as the helical orders in Figs. \ref{fig-A5}(a) and (b).  However, this expectation is contradictory to the snapshots in  Figs. \ref{fig-A5}(a) and (b), where the helical orders along the arrow directions are stable. This contradiction implies again that FMI anisotropy, i.e., the resulting direction of $\vec{\tau}$ by the lattice deformation, in model 1 implemented by FG modeling is not suitable. 

\section*{References}


\end{document}